\documentclass[a4paper]{scrartcl}
\usepackage{fix-cm}
\usepackage[T1]{fontenc}
\usepackage{lmodern}

\DeclareOldFontCommand{\it}{\normalfont\itshape}{\mathit}
\DeclareOldFontCommand{\bf}{\normalfont\bfseries}{\mathbf}
\DeclareOldFontCommand{\rm}{\normalfont\rmfamily}{\mathrm}
\DeclareOldFontCommand{\sf}{\normalfont\sffamily}{\mathsf}
\DeclareOldFontCommand{\tt}{\normalfont\ttfamily}{\mathtt}
\DeclareOldFontCommand{\sc}{\normalfont\scshape}{\@nomath\sc}

\usepackage{epsfig}
\usepackage{amssymb}
\usepackage{graphicx,color}
\usepackage{amsmath}
\usepackage{array}
\usepackage[center]{subfigure}
\usepackage{slashed}
\usepackage{bbm}
\usepackage{mathrsfs}
\usepackage{multirow}
\usepackage{hyperref}
\usepackage{xcolor}
\usepackage{amsmath}
\usepackage{slashed}
\usepackage{comment}

\usepackage{booktabs}
\usepackage{siunitx} 
\usepackage[utf8]{inputenc}

\usepackage[normalem]{ulem}

\newcommand{\bra}[1]{\left\langle #1 \right|}
\newcommand{\ket}[1]{\left| #1 \right\rangle}

\newcommand{\be}{\begin{equation}}
\newcommand{\ee}{\end{equation}}
\newcommand{\ba}{\begin{eqnarray}}
\newcommand{\ea}{\end{eqnarray}}

\definecolor{darkermagenta}{rgb}{0.7, 0, 0.7}
\definecolor{darkgreen}{RGB}{0,175,10}
\definecolor{AK}{rgb}{0,0,1.0}

\begin{document}

 \begin{titlepage}
\pagestyle{empty}
\begin{flushright}
SI-HEP-2026-03\\
P3H-26-012
\end{flushright}
\vfill
 \begin{center}
 {\Large \boldmath
 \textbf{
Light-cone sum rules 
with $B$-meson
distribution amplitudes
for the $B\to p$ form factors \\in  
$B$-mesogenesis models 
 }\\[.5cm]}
 
{\large Aritra Biswas, Alexander Khodjamirian and Ali Mohamed }
\vspace*{0.4cm}

\textsl{%
Center for Particle Physics Siegen (CPPS), Theoretische Physik 1\,,\\
Universit\"at Siegen, D-57068 Siegen, Germany \\[3mm]
}

\begin{abstract}

New decay modes of $B$-meson  into 
a baryon and invisible dark antibaryon $\Psi$
are among the most distinctive signatures of the $B$-mesogenesis scenario.  
We concentrate on the proton mode and consider  
two versions of the underlying interaction of $\Psi$ with quarks, the so-called models
$(d)$ and $(b)$. To estimate the width of the $B^+\to p \Psi $  decay, we obtain  the $B^+\to p$ transition
form factors, applying QCD light-cone sum rules (LCSRs) with
$B$-meson distribution amplitudes with an accuracy up to twist-5, while interpolating the  proton with a current. 
This method is independent of the 
previously applied one, which was based on the nucleon  distribution amplitudes. We estimate the partial width of the  $B^+\to p \Psi $  decay as a function of the dark antibaryon  mass. Furthermore, we  use the ratio of this width to the inclusive $B\to X_N \Psi $ width, the latter predicted
earlier using the heavy quark expansion method.
This ratio, which is independent of the effective coupling,
when combined with the minimal 
inclusive branching fraction of $O(10^{-4})$, necessary for  the 
feasibility of $B$-mesogenesis, 
yields lower limits on the $B^+\to p \Psi$ branching fraction.
 We confront these limits with the 
most recent upper bounds obtained  from BaBar and Belle/Belle II searches for the decays of $B^+$-meson into a proton and missing energy.
The comparison indicates that experimental upper bounds on the branching fraction of 
$B\to p \Psi $ at the level of $10^{-8}-10^{-7}$ are needed for a decisive probe of this invisible mode of $B$ decays. 

\end{abstract}
\end{center}

 \vfill
 \end{titlepage}

\section{Introduction}

The $B$-mesogenesis scenario  suggested in \cite{Elor:2018twp,Alonso-Alvarez:2021qfd} (see also \cite{Nelson:2019fln,Elahi:2021jia,Alonso-Alvarez:2021oaj}) has caused a lot of interest,
offering a possibility to 
simultaneously
solve  two fundamental problems:  the baryon 
asymmetry and the dark matter abundance  in the Universe. Moreover, the $B$-meson decay modes into 
a light baryon and dark antibaryon $\Psi$ predicted in this scenario are accessible in current experimental searches for invisible $B$ decays. The most recent 
search \cite{Belle-II:2026tyb} for these decays by Belle II collaboration (using the data of Belle experiment) yields upper limits reaching  $O(10^{-6})$. Earlier 
measurements
with less restrictive results
included searches for 
$B^+\to p \Psi $, $B^0\to\Lambda \Psi$
and $B^+\to\Lambda^+_c \Psi$~
at the BaBar experiment (see 
\cite{BaBar:2023dtq}, ~\cite{BaBar:2023rer}, 
and \cite{BaBar:2024qqx}, respectively) and 
the  search for $B^0\to\Psi\Lambda$
at Belle~\cite{Belle:2021gmc}. 

On the theory side, the main task is 
to estimate the branching fraction of a $B$ decay into a light baryon and $\Psi$ for a given effective  interaction of quarks with $\Psi$. 
In this paper, we focus on the decay mode $B^+\to p \Psi$, which has the largest available phase space and therefore allows for the largest accessible values of the $\Psi$ mass.
We consider the two versions of a possible  $\Psi$-triquark interaction proposed in \cite{Elor:2018twp,Alonso-Alvarez:2021qfd}. The same versions were considered  
in \cite{Khodjamirian:2022vta},  
where they were denoted  as the $B$-mesogenesis models $(d)$ and $(b)$. In the first (second) model, the $d$-quark ($b$-quark)  is directly coupled to  $\Psi$ and a heavy mediator. The effective pointlike $\Psi$-triquark   interaction that leads
to the $B^+\to p\Psi$ decay is then formed by the coupling of the $d\Psi$ 
($b\Psi$) pair via heavy mediator to the $ub$ ($ud$) diquark.

Given the effective operators of the $\Psi$-triquark  interaction, 
one still needs reliable estimates of 
the $B\to$ baryon transitions form factors.
To this end, the QCD-based method of light-cone sum rules (LCSRs) was used in \cite{Khodjamirian:2022vta} (see also \cite{Elor:2022jxy,Boushmelev:2023huu}). The sum rules 
for $B\to p$ form factors were obtained, 
applying a vacuum-to-proton correlation function, 
where the effective operator of the new interaction was combined 
with the $B$-meson interpolating current.
In the resulting LCSRs, nonperturbative input included the 
 light-cone distribution amplitudes (DAs) of the proton, i.e. of the nucleon in the adopted 
isospin symmetry limit. 
In the latest analysis  of these LCSRs \cite{Boushmelev:2023huu}, the accuracy 
of the operator product expansion (OPE) was upgraded from the initial twist-3 level to include all terms up to twist-6.
As a result, large higher-twist  contributions were revealed, signaling a slow convergence of twist expansion. Note also that the 
nucleon DAs used in the LCSRs
depend on multiple parameters, determining their normalizations and non-asymptotic terms, and some of them
still have substantial uncertainties.

All that calls for an independent estimate of the $B\to p$  form factors,
which is the main goal of this paper.
Here we apply a different version of LCSRs, where the roles of 
the final proton and initial $B$-meson in the underlying correlation
function  are inverted,  as compared to the sum rules based on nucleon DAs.
We use a vacuum-to-$B$ correlation function of two local operators,
one of them being the three-quark current interpolating the proton and the 
other one is the operator of the effective $\Psi$-triquark interaction.
    Accordingly, the  nonperturbative input in the OPE for this correlation function is represented by a set of $B$-meson DAs~\cite{Grozin:1996pq} defined  in heavy quark effective theory (HQET). The 
 twist expansion of these DAs was elaborated  in \cite{Braun:2017liq}.
Previously, this method was applied to various
$B$-meson transitions to light mesons (see e.g.
\cite{Khodjamirian:2005ea,Khodjamirian:2006st,Gubernari:2018wyi}) as well as to $B\to D^{(*)}$ form factors \cite{Faller:2008tr}. For a recent review and comparison with other versions of LCSRs see also \cite{Khodjamirian:2023wol}.

The main advantage of the method based on $B$-meson DAs is a straightforward possibility to  access 
various $B\to$ baryon form factors,   replacing
the proton interpolation current by any other three-quark 
current with a light or charmed baryon quantum numbers, 
while the main nonperturbative input  remains the same. This allows one  to obtain 
the form factors  corresponding to different flavour structures of the effective three-quark-$\Psi$ interaction 
in the mesogenesis models.
In this paper, we confine ourselves to the proton mode; the others will be considered elsewhere. 

Our main results are the $B\to p$ form factors 
obtained from LCSRs for both models $(d)$ and $(b)$ in the region $q^2<0$ of  spacelike four-momentum transfer $q$. These form factors are then extrapolated to
positive values of $q^2=m_\Psi^2$ in the whole region of $\Psi$ masses allowed by  the mesogenesis scenario.
We finally predict the 
$B^+\to p \Psi$ decay branching fraction as a function of the $\Psi$ mass. 
For these estimates we adopt, for definiteness, the 
$\Psi$-triquark couplings  equal to their upper limits inferred
in~\cite{Alonso-Alvarez:2021qfd} from the LHC 
searches for new heavy particles.

In addition, we use the results of   \cite{Lenz:2024rwi,Mohamed:2025zgx} (see also \cite{Shi:2023riy}), where, applying the 
heavy quark expansion method, the width of inclusive decay 
$B\to X_N\Psi$  was obtained in the same models $(d)$ and $(b)$. Here $X_N$ denotes the sum over all kinematically accessible hadronic states 
generated by fragmentation of the partonic final state, 
formed by the diquark $ud$ of
the effective $\Psi$-triquark interaction and by the spectator $u$-quark of $B^+$-meson.  
Dividing our new  estimates of  $B^+\to p\Psi$
branching fractions to the  corresponding inclusive widths,
we obtain their ratios
in which the unknown $\Psi$-triquark coupling is canceled.
The inclusive invisible $B$ decays  cannot be experimentally
searched for. Importantly, however, there is a  theory-motivated
 estimate \cite{Elor:2018twp,Alonso-Alvarez:2021qfd} of  the minimal inclusive width, necessary for the realization of the  $B$-mesogenesis scenario. We use that bound, together with the estimated exclusive-to-inclusive ratios and obtain
the corresponding lower limits on the 
exclusive $B^+\to p\Psi$ branching fraction, which only depend on the $\Psi$ mass
and on the model of $\Psi$-triquark interaction. The  limits we obtain are then confronted with the   upper bounds measured by  BaBar and Belle/Belle II. 

The paper is organized as follows: in Section~\ref{sect:coupl} we specify the effective interactions emerging in the 
considered models of $B$-mesogenesis. In Section~\ref{sect:lcsr}  we derive the new LCSRs for the $B\to p$ form factors and in Section~\ref{sect:ope} we present and discuss the results of light-cone OPE in terms of $B$-meson DAs.
Section~\ref{sect:num} contains our numerical results. After specifying the input parameters, we present  
the form factors calculated from LCSRs at spacelike momentum transfer. Employing $z$-expansion, these form factors are continued to the physical values of $\Psi$ mass, yielding our results for the $B\to p\Psi$ width.
In Section~\ref{sect:excl-incl} we use these results to obtain the ratios of exclusive and inclusive widths.
Our concluding discussion is in Section~\ref{sect:concl}. Appendix \ref{sect:appBDA} contains the details of $B$-meson DAs and in Appendix \ref{app:disp} we present a derivation
of the dispersion representation for logarithmic terms in the OPE.

\section{The effective couplings and $B^+\to p \Psi$ decay amplitude}
\label{sect:coupl}
For the effective Hamiltonian of the $\Psi$-triquark interaction generating the $B^+\to p \Psi$ decay and its charge conjugate mode, we use  the same compact form as in 
\cite{Khodjamirian:2022vta}:
 \ba
 {\cal H}_{(-1/3)} \, =
 - \; G_{(d)}\overline{{\cal O}}_{(d)} \Psi^c 
  -\:
 G^*_{(d)}\bar{\Psi}^c{\cal O}_{(d)}+
 \{d\leftrightarrow b \}\,,~~~~
 \label{eq:Heff2}
 \ea
 where $G_{(d)}$ is the effective four-fermion coupling,
 inversely proportional to the squared mass  of the heavy mediator particle in the 
 $B$-mesogenesis scenario (see \cite{Elor:2018twp,Alonso-Alvarez:2021qfd} for details).  
 The local three-quark operator and its conjugate in (\ref{eq:Heff2}) are 
 \be
 \overline{{\cal O}}_{(d)}= i \epsilon_{ijk}\left(\bar{u}^i_{R} b_R^{c\,j}\right)\bar{d}^{\,k}_R, ~~
 {\cal O}_{(d)}= i \epsilon_{ijk}d^{\,i}_R \left(\bar{b}^{c\,j}_R u^{\,k}_R\right)\,. 
 \label{eq:Od}
 \ee
In the above equation, the  index $c$ ($R$) indicates  charge conjugated (right-handed) fields, so that
 $q_R=\frac12(1+\gamma_5)q$; and $i,j,k$ are the colour indices. 
 The index $(d)$
 indicates  the terms in the effective Hamiltonian (\ref{eq:Heff2})
 originating from the $d$-quark coupling 
 to the $\Psi$ and  heavy mediator fields, that is  model $(d)$. Replacing $d\leftrightarrow b $ yields 
model $(b)$, which corresponds to the terms in (\ref{eq:Heff2}) 
 with the effective coupling $G_{(b)}$
and with the three-quark operators
 \be
 \overline{{\cal O}}_{(b)}= i \epsilon_{ijk}\left(\bar{u}^i_{R} d_R^{\,c\,j}\right)\bar{b}^k_R, ~~
 {\cal O}_{(b)}= i \epsilon_{ijk}b^{\,i}_R \left(\bar{d}^{\,c\,j}_R u^{\,k}_R\right) \; .
 \label{eq:Ob}
 \ee

Considering as our main study case model $(d)$,
we write down the amplitude of the $B^+\to p \Psi$ decay 
generated by the effective operator $\overline{{\cal  O}}_{(d)}$:
\ba
{\cal A}_{(d)}(B^+\to p \Psi)&=& G_{(d)}
\langle p(P)| \overline{{\cal  O}}_{(d)} |B^+(P+q)\rangle  u^c_{\Psi}(q)\,,
\label{eq:ampld}
\ea
where $u^c_\Psi$ is the (charge conjugated) bispinor of the $\Psi$ field,
and  the following on-shell conditions are fulfilled: $P^2=m_p^2$\,, $(P+q)^2=m_B^2$,
$q^2=m_\Psi^2$.

The hadronic 
matrix element of the three-quark operator in (\ref{eq:ampld}) can be parameterized, 
in the most general form, in terms of four independent $B\to p$ form factors: 
\ba
\langle p(P)| \overline{{\cal  O}}_{(d)} |B^+(P+q)\rangle =
F^{(d)}_{B\to p_R}(q^2)\bar{u}_{pR}(P)+
F^{(d)}_{B\to p_L}(q^2)\bar{u}_{pL}(P)
\nonumber\\
+
\widetilde{F}^{(d)}_{B\to p_R}(q^2)\bar{u}_{pR}(P) \frac{\slashed{q}}{m_p}+ \widetilde{F}^{(d)}_{B\to p_L}(q^2)\bar{u}_{pL}(P) \frac{\slashed{q}}{m_p}\,,
\label{eq:ff}
\ea 
where $u_{p\,R(L)}(P)\equiv \gamma_{R(L)}u_{p}(P)$ and 
$\bar{u}_{p\,R(L)}\equiv  \bar{u}_p(P)\gamma_{L(R)} $ are
the bispinor of the right-handed (left-handed) proton and its Dirac-conjugate. Hereafter we use  a compact notation:
$$ \gamma_R\equiv \frac12(1+\gamma_5), ~~
\gamma_L\equiv \frac12(1-\gamma_5)\,.$$
The expansion in terms of four form factors in 
(\ref{eq:ff}) corresponds  to the  most general Lorentz/Dirac decomposition in 
four independent kinematical structures that can be
 constructed using a bispinor and two independent four-momenta $P$ and $q$,
 and is independent of the particular chiral structure of the operator $\overline{{\cal  O}}_{(d)}$.
Note that the structures $\bar{u}_{p\,L,R}(P)\slashed{P}$ are
reduced to $\bar{u}_{p\,L,R}(P)$ 
after applying  the Dirac equation for bispinors. 
For model $(b)$, the same decay amplitude and its decomposition
in terms of four form factors are simply obtained from (\ref{eq:ampld}) and 
(\ref{eq:ff}), respectively, by the replacement $(d)\to (b)$.

\section{LCSR with $B$-meson distribution amplitudes}
\label{sect:lcsr}

Here we derive the LCSRs for the $B\to p$ form factors. 
For definiteness, we first consider model $(d)$ with the three-quark  operator 
$\overline{\cal O}_{(d)}$
specified in (\ref{eq:Od}).
The key element of our method is 
the correlation function
\be
{\cal F}^{(d)}(P,q)=i\int d^4x\ e^{iP\cdot
x}\bra{0}T\left\{\eta_p(x),{\overline{{\cal O}}_{(d)}} (0)\right\}\ket {B^+(P+q)} \,.
\label{eq:corr}
\ee
It contains the time-ordered product of 
${\overline{{\cal O}}_{(d)}}$ and 
the proton interpolating current $\eta_p$, for which we choose the Ioffe current \cite{Ioffe:1981kw}, known to be an optimal choice  for other QCD sum rules in the nucleon 
channel: 
\be
 \eta_p(x)= \epsilon_{ijk} \big[ u^{Ti}(x) C \gamma^{\mu} u^j(x)\big] \gamma_{5} \gamma_{\mu} d^k(x),
 \label{eq:IoffeCurrent}
\ee 
where the color indices are shown explicitly and $C$ is the charge conjugation matrix. In our convention $C=\gamma^2\gamma^0$.  
The bilocal operator product in (\ref{eq:corr}) is 
sandwiched between the  on-shell $B$-meson state and the vacuum.

The correlation function (\ref{eq:corr}) is by construction a Dirac matrix, which is Lorentz-contracted  with the two independent four-momenta $P$ and $q$. Its generic
decomposition in the 
kinematical structures reads:
\begin{align}
{\cal F}^{(d)}(P,q) &= 
\Big[{\cal F}^{(d)}_{1}(P^2,q^2)
+ {\cal F}^{(d)}_{2}(P^2,q^2)\frac{\slashed{P}}{m_B}
+ {\cal F}^{(d)}_{3}(P^2,q^2)\frac{\slashed{q}}{m_B}
+ {\cal F}^{(d)}_{4}(P^2,q^2)\frac{\slashed{P}\slashed{q}}{m_B^2}
\Big]\gamma_L
\nonumber\\
\quad &+
\Big[{\cal F}^{(d)}_{5}(P^2,q^2)
+ {\cal F}^{(d)}_{6}(P^2,q^2)\frac{\slashed{P}}{m_B}
+ {\cal F}^{(d)}_{7}(P^2,q^2)\frac{\slashed{q}}{m_B}
+ {\cal F}^{(d)}_{8}(P^2,q^2)\frac{\slashed{P}\slashed{q}}{m_B^2}
\Big]\gamma_R \, ,
\label{eq:corrdecomp}
\end{align}
where ${\cal F}^{(d)}_{1,...8}$ are Lorentz-invariant
amplitudes depending on the invariant variables $P^2$ and $q^2$.

To access the $B^+\!\to\!p$ form factors,
we employ the unitarity relation, expressing the imaginary part of the correlation function
(\ref{eq:corr})
in the timelike region $P^2>0$ of the variable $P^2$, 
via the sum of all possible intermediate hadronic states 
with the quantum numbers of the interpolating current.
We have:
\ba
\frac{1}{\pi}\, \text{Im}\,{\cal F}^{(d)}(P,q)= \delta(P^{2}-m_p^2) \bra{0} \eta_{p}\overline{ \ket{p^+(P)} \bra{p^+(P)}} {\bar{{\cal O}}_{(d)}} \ket{B^+(P+q)} 
\nonumber\\
+\delta(P^{2}-m_{N^*}^2) \bra{0} \eta_{p} \overline{\ket{N^{*+}(P)} \bra{N^{*+}(P)}} {\bar{{\cal O}}_{(d)}} \ket{B^+(P+q)} 
+\rho_h^{(d)}(P,q)\,,
\label{eq:unitary}
\ea
where the overlines indicate summing over baryon polarizations.
Above, we take into account that the current (\ref{eq:IoffeCurrent}) interpolates  baryons with both positive and negative $P$-parities. 
Hence, in addition to the proton contribution,
we also isolate the contribution of 
$N^*\equiv N(1535)$, 
the lowest nucleon resonance with spin-parity
$J^P=1/2^-$, see 
\cite{ParticleDataGroup:2024cfk}.  
Resonances and continuum states with quantum numbers $J^P=1/2^+$ ($J^P=1/2^-$) and heavier than the  proton (the $N^*$)
are, by default, included in the spectral density $\rho_h^{(d)}(P,q)$. 

We define the hadronic matrix elements of interpolation 
current via the 
proton and $N^*$ decay constants:
\be
 \bra{0} \eta_{p} \ket{p(P)}=\lambda_p m_p u_p(P)\,,~~
 \bra{0} \eta_{p} \ket{N^{*+}(P)} = 
 \lambda_{N^*} m_{N^*}\gamma_5 u_{N^*}(P)\,,
 \label{eq:deconst}
\ee
where $u_p$ ($u_{N^*}$) is the Dirac bispinor for the proton ($N^*$) state
with four-momentum $P$. The decay constants $\lambda_{p,N^*}$ are 
obtained from two-point QCD sum rules or 
computed in lattice QCD.

Apart from the definition 
(\ref{eq:ff}), 
we also need an analogous expansion  of the $B\to N^*$ hadronic matrix element in terms of corresponding form factors:
\ba
\langle N^*(P)| \bar{{\cal  O}}_{(d)} |B^+(P+q)\rangle =
F^{(d)}_{B\to N^*_R}(q^2)\bar{u}_{N^*R}(P)\gamma_5+
F^{(d)}_{B\to N^*_L}(q^2)\bar{u}_{N^*L}(P)\gamma_5\nonumber\\
+
\widetilde{F}^{(d)}_{B\to N^*_R}(q^2)\bar{u}_{N^*R}(P)\gamma_5\frac{\slashed{q}}{m_{N^*}}+ \widetilde{F}^{(d)}_{B\to N^*_L}(q^2)\bar{u}_{N*L}(P) \gamma_5\frac{\slashed{q}}{m_{N^*}}\,,
\label{eq:ffNst}
\ea 
Substituting (\ref{eq:ff}), (\ref{eq:deconst}) and (\ref{eq:ffNst})
in (\ref{eq:unitary}) and summing over polarizations, we obtain
\begin{align}
\frac{1}{\pi}\text{Im}\,{\cal F}^{(d)}(P,q)=  \delta(P^{2}-m_p^2) \, \lambda_p m_p \Bigg\{
F^{(d)}_{B\to p_R}(q^2)(\slashed{P}+m_p) \gamma_L+
F^{(d)}_{B\to p_L}(q^2)(\slashed{P}+m_p) \gamma_R
\nonumber\\
+\widetilde{F}^{(d)}_{B\to p_R}(q^2)(\slashed{P}+m_p) \frac{\slashed{q}}{m_p}\gamma_R+ \widetilde{F}^{(d)}_{B\to p_L}(q^2)(\slashed{P}+m_p)  \frac{\slashed{q}}{m_p}\gamma_L\, \Bigg\}
\nonumber\\
-\delta(P^{2}-m_{N^*}^2) \, \lambda_{N^*} m_{N^*}\Bigg\{
F^{(d)}_{B\to N^*_R}(q^2)(\slashed{P}-m_{N^*}) \gamma_L+
F^{(d)}_{B\to N^*_L}(q^2)(\slashed{P}-m_{N^*}) \gamma_R
\nonumber\\
+\widetilde{F}^{(d)}_{B\to N^*_R}(q^2)(\slashed{P}-m_{N^*}) \frac{\slashed{q}}{m_{N^*}}\gamma_R  + 
\widetilde{F}^{(d)}_{B\to N^*_L}(q^2)(\slashed{P}-m_{N^*}) \frac{\slashed{q}}{m_{N^*}} \gamma_L\, \Bigg\}
+\rho^{(d)}(P,q)\,,
\label{eq:unitary2}
\end{align}
\noindent
where it is easy to identify the kinematical structures 
multiplying various form factors with the ones in the 
decomposition (\ref{eq:corrdecomp}). In general, 
the spectral density 
$\rho^{(d)}(P,q)$ is also decomposed in eight
Lorentz-invariant functions:
\ba
\rho^{(d)}(P,q)=\big[\rho^{(d)}_{1}(P^2,q^2)  + 
\rho^{(d)}_{2}(P^2,q^2)\frac{\slashed{P}}{m_B}+
\rho^{(d)}_{3}(P^2,q^2)\frac{\slashed{q}}{m_B}+
\rho^{(d)}_{4}(P^2,q^2) \frac{\slashed{P}\,\slashed{q}}{m_B^2}\big] 
\gamma_L 
\nonumber\\
+\big[\rho^{(d)}_{5}(P^2,q^2)  + 
\rho^{(d)}_{6}(P^2,q^2)\frac{\slashed{P}}{m_B}+
\rho^{(d)}_{7}(P^2,q^2)\frac{\slashed{q}}{m_B}+
\rho^{(d)}_{8}(P^2,q^2) \frac{\slashed{P}\,\slashed{q}}{m_B^2}\big]
\gamma_R\,.
\label{eq:rhodecomp}
\ea

The imaginary part (\ref{eq:unitary2}) of the correlation function enters the dispersion relation 
for this function in the variable $P^2$ at fixed $q^2$ 
\footnote{In (\ref{eq:had-disp}),
we ignore possible subtraction terms, since later they will be eliminated by the Borel transform.}:
\be
{\cal F}^{(d)}(P,q)=\frac{1}{\pi} 
\int\limits_{m_p^2}^{\infty} 
ds \frac{\text{Im}\,{\cal F}^{(d)}(P,q)}{s-P^2}\,.
\label{eq:had-disp}
\ee

We substitute the decompositions 
(\ref{eq:corrdecomp}) and (\ref{eq:unitary2}),(\ref{eq:rhodecomp}) 
into  the left-hand side  and, respectively, right-hand side of the above relation.
Equating the coefficients at each independent 
kinematical structure on both sides, we obtain separate dispersion relations for the
eight independent invariant amplitudes. For the first four amplitudes
these relations read:
\ba
{\cal F}^{(d)}_{1}(P^2,q^2) = \frac{\lambda_p m_p^2}{m^2_p - P^2} 
F^{(d)}_{B\to p_R}(q^2)+
\frac{\lambda_{N^*} m_{N^*}^2}{m^2_{N^*} - P^2} F^{(d)}_{B\to N^*_R}(q^2)
+\int\limits_{s_h} ^{\infty} ds\,\frac{\rho_1^{(d)}(s,q^2)}{s-P^2}, \label{eq:Fd1disp}
\ea
\ba
{\cal F}^{(d)}_{2}(P^2,q^2) = \frac{\lambda_p m_p m_B}{m^2_p - P^2} 
F^{(d)}_{B\to p_R}(q^2)-
\frac{\lambda_{N^*} m_{N^*}m_B}{m^2_{N^*} - P^2} F^{(d)}_{B\to N^*_R}(q^2)
+\int\limits_{s_h} ^{\infty} ds\,\frac{\rho_2^{(d)}(s,q^2)}{s-P^2}, \label{eq:Fd2disp}
\ea
\ba
{\cal F}^{(d)}_{3}(P^2,q^2) = \frac{\lambda_p m_p m_B}{m^2_p - P^2} 
\widetilde{F}^{(d)}_{B\to p_L}(q^2)+
\frac{\lambda_{N^*} m_{N^*}m_B}{m^2_{N^*} - P^2} \widetilde{F}^{(d)}_{B\to N^*_L}(q^2)
+\int\limits_{s_h} ^{\infty} ds\,\frac{\rho_3^{(d)}(s,q^2)}{s-P^2}, \label{eq:Fd3disp}
\ea
\ba
{\cal F}^{(d)}_{4}(P^2,q^2) = \frac{\lambda_p m_B^2}{m^2_p - P^2} 
\widetilde{F}^{(d)}_{B\to p_L}(q^2)-
\frac{\lambda_{N^*} m_B^2}{m^2_{N^*} - P^2} \widetilde{F}^{(d)}_{B\to N^*_L}(q^2)
+\int\limits_{s_h} ^{\infty} ds\,\frac{\rho_4^{(d)}(s,q^2)}{s-P^2}\,. \label{eq:Fd4disp}
\ea
The four remaining relations for the invariant amplitudes
${\cal F}^{(2)}_{5,6,7,8}$
are obtained from the above four equations by the following replacements of indices: 
in  Eq.~(\ref{eq:Fd1disp}): $1\to 5$, $R\to L$; 
in  Eq.~(\ref{eq:Fd2disp}): $2\to 6$, $R\to L$ ;
in  Eq.~(\ref{eq:Fd3disp}): $3\to 7$, $L\to R$ ;
in  Eq.~(\ref{eq:Fd4disp}): $4\to 8$, $L\to R$. 
The lower limit $s_h$ in the integrals over heavier 
than proton continuum states is the same in all dispersion relations: $s_h=(m_p+m_{\pi^0})^2$. It corresponds to the lowest possible threshold of 
the positively charged pion-nucleon state with $J=1/2$.

Below, we will calculate
the correlation function
(\ref{eq:corr}), employing  the light-cone OPE in terms of the $B$-meson DAs. 
The OPE is applicable at 
spacelike $P^2$, such that $|P^2|\gg\Lambda_{QCD}^2$, far below the position of the proton pole at 
$P^2=m_p^2$. This condition will be satisfied after Borel transform,  choosing the Borel mass
parameter sufficiently large. Simultaneously, the variable $q^2$ has to be chosen also far from respective 
hadronic states in the $bud$ channel with the quantum numbers 
of $\Lambda_b$. Keeping $q^2<0$ is then sufficient. Our
results for all invariant amplitudes 
in the region of OPE validity
will be cast in a convenient form of a  dispersion relations in the $P^2$ variable :
\be
{\cal F}^{(d)_{OPE}}_n(P^2,q^2)=\frac{1}{\pi}\int\limits_{0} ^\infty ds\frac{\mbox{Im}{\cal F}_n^{(d)_{OPE}}(s,q^2)}{s-P^2}\,,~~~(n=1,...,8)
\label{eq:OPEdsip}
\ee
where the lower limit corresponds to the adopted 
approximation of  massless $u,d$ quarks
in the OPE of the correlation function.
The representation (\ref{eq:OPEdsip}) allows us to employ
the usual assumption of (semi-global) quark-hadron duality:
\be
\int\limits_{s_h} ^\infty ds\frac{\rho_n^{(d)}(s,q^2)}{s-P^2}=
\frac{1}{\pi}\int\limits_{s_{0(n)}} ^\infty ds\frac{\mbox{Im}{\cal F}_n^{(d)OPE}(s,q^2)}{s-P^2}\,,
~~~(n=1,...,8)
\label{eq:dual}
\ee
introducing the effective threshold  $s_{0(n)}$, which is in general a specific parameter 
for each invariant amplitude ${\cal F}^{(d)}_n$, hence the index $(n)$.

We substitute~(\ref{eq:OPEdsip}) into the left-hand side and, respectively, (\ref{eq:dual}) into the right-hand side of each dispersion relation in~(\ref{eq:Fd1disp})--(\ref{eq:Fd4disp}) and their counterparts for the invariant amplitudes ${\cal F}^{(2)}_{5-8}$.
After that,  we subtract from both sides the integrals over OPE spectral densities from $s_{0(n)}$ to $\infty$.
Performing then the Borel transform $P^2\to M^2$, we obtain a set of eight LCSRs containing  all 
$B\to p$ and $B\to N^*$ form factors. 
The first four sum rules obtained from (\ref{eq:Fd1disp})-(\ref{eq:Fd4disp})
are: 
\ba
\lambda_p m_p^2\, e^{-\frac{m_p^2}{M^2}} F^{(d)}_{B\to p_R}(q^2)\!&+&\!
\lambda_{N^*} m_{N^*}^2\,e^{-\frac{m^2_{N^*}}{M^2}}
F^{(d)}_{B\to N^*_R}(q^2)
\nonumber\\
&=&\frac{1}{\pi}\!\int\limits_{0} ^{s_{0(1)}}\!\!ds\,e^{-\frac{s}{M^2}}
\,\text{Im} {\cal F}_1^{(d)_{OPE}}(s,q^2)\,, 
\label{eq:lcsr1}
\ea
\ba
\lambda_p m_pm_B\, e^{-\frac{m_p^2}{M^2}} F^{(d)}_{B\to p_R}(q^2) 
&-& \lambda_{N^*} m_{N^*}m_B\,e^{-\frac{m^2_{N^*}}{M^2}}
F^{(d)}_{B\to N^*_R}(q^2)
\nonumber\\
&=&\frac{1}{\pi}\int\limits_{0} ^{s_{0(2)}}\!\! ds\,e^{-\frac{s}{M^2}}
\,\text{Im} {\cal F}_2^{(d)_{OPE}}(s,q^2)\,, 
\label{eq:lcsr2}
\ea
\ba
\lambda_p m_pm_B\, e^{-\frac{m_p^2}{M^2}} \widetilde{F}^{(d)}_{B\to p_L}(q^2)\!&+&\!
\lambda_{N^*} m_{N^*}m_B\,e^{-\frac{m^2_{N^*}}{M^2}}
\widetilde{F}^{(d)}_{B\to N^*_L}(q^2)
\nonumber\\
&=&\frac{1}{\pi}\int\limits_{0} ^{s_{0(3)}}\!\! ds\,e^{-\frac{s}{M^2}}
\,\text{Im} {\cal F}_3^{(d)_{OPE}}(s,q^2)\,, 
\label{eq:lcsr3}
\ea
\ba
\lambda_p m_B^2\, e^{-\frac{m_p^2}{M^2}} \widetilde{F}^{(d)}_{B\to p_L}(q^2)\!&-&\!
\lambda_{N^*} m_B^2\,e^{-\frac{m^2_{N^*}}{M^2}}
\widetilde{F}^{(d)}_{B\to N^*_L}(q^2)
\nonumber\\
&=&\frac{1}{\pi}\int\limits_{0} ^{s_{0(4)}}\!\! ds\,e^{-\frac{s}{M^2}}
\,\text{Im} {\cal F}_4^{(d)_{OPE}}(s,q^2)\,. 
\label{eq:lcsr4}
\ea
It is easy to obtain analogous sum rules for the amplitudes ${\cal F}^{(d)}_{5-8}$,   applying to the above equations the replacements described after (\ref{eq:Fd4disp}). 
The analogous sum rules for model $(b)$ have the same 
expressions, as the ones for model $(d)$, only the 
index $(d)\to (b)$ should be replaced at the form factors and OPE amplitudes.

The LCSRs derived above 
can now be used to estimate the $B\to p$ 
and, in principle, also the $B\to N^*$ form factors, provided the OPE amplitudes entering these sum rules via their spectral densities are obtained with a certain accuracy.
For the channel of nucleon current,
there is however an ambiguity in the separation of hadronic states 
into the two lowest resonances and a duality-approximated sum over excited  and continuum states. A prominent and very broad Roper resonance $N(1440)$ with the same spin-parity 
$J^P=1/2^+$ as
the nucleon  is located 
below $N^*(1535)$ 
\cite{ParticleDataGroup:2024cfk}. Hence, for completeness, 
in each of the LCSRs (\ref{eq:lcsr1})--(\ref{eq:lcsr4})
a $N(1440)$ 
contribution should  be then added  as a separate term proportional to 
a $B\to N(1440)$ form factor. The expressions for these terms 
are simply obtained by replacing in the proton term 
$p\to N(1440)$.
A separate issue, which we do not discuss here, concerns accounting for the large total widths of $N^*$ and $N(1440)$ in the resonance-pole terms.

Disentangling the $B\to p$ form factors from sum rules that contain two 
additional resonance terms would require a dedicated fit 
procedure, for which the knowledge of all three decay constants
$\lambda_p$, $\lambda_{N^*}$ and $\lambda_{N(1440)}$ is  necessary. 
However, we are unaware of any reliable estimate of the  $N(1440)$ decay constant. Hence, in our numerical analysis
performed below, we leave such a separation procedure beyond our scope. Instead,
we  employ a simplified form of the hadronic part, following previous analyses of LCSRs for
the nucleon electromagnetic form factors (see e.g., \cite{Anikin:2013aka,
Braun:2001tj}). In those sum rules the same proton (nucleon) interpolating current was used and the nucleon resonances $N(1440)$ and $N^*(1535)$ were incorporated into the spectral density $\rho_n^{(d)}(s)$ via the duality approximation~(\ref{eq:dual}), assuming a uniform effective threshold, which we also adopt here, such that 
$s_{0(n)}=s_0$.

We now turn to 
 the OPE of the correlation function 
(\ref{eq:corr}), aiming at 
the OPE spectral density that enters LCSRs.

\section{OPE for the correlation function}
\label{sect:ope}
\begin{figure}[t]
\begin{center}
\includegraphics[scale=0.5]{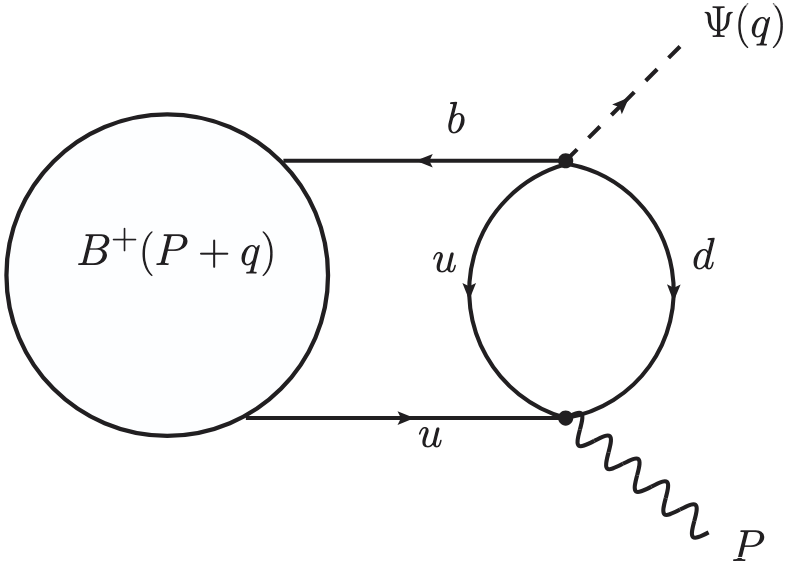}
\end{center}
\caption{Diagram corresponding to the 
contribution of two-particle $B$-meson DAs to the correlation function (\ref{eq:corr}). The wavy (dashed) line represents the interpolating current of the proton (the dark antibaryon).}
\label{fig:diag}
\end{figure}
\subsection{Contribution of two-particle DAs}

To describe  the calculation of the 
correlation function in detail, we take, for 
definiteness, model $(d)$ and substitute in
(\ref{eq:corr}) the expression 
(\ref{eq:Od})
for the operator
${\overline{{\cal O}}_{(d)}}$
in terms of quark fields. Contracting the $u$ and $d$ fields 
in the free massless-quark propagators, we encounter 
the leading-order (LO) diagram depicted in Fig.~\ref{fig:diag}.
At this level, 
due to the presence of three-quark currents, the diagram already contains a diquark loop \footnote{ Note that in
 previous applications
of the LCSRs with $B$-meson DAs to the $B\to $ meson form factors , the 
hard-scattering part of 
LO diagrams represents 
a single virtual quark line connecting two vertices of currents.}.

The remaining 
matrix element in the expression for the
LO diagram is the product of $\bar{b}$- and $u$-quark fields 
at points 0 and $x$, respectively,  sandwiched between 
the $B^+$-meson and vacuum states. For that matrix element, we employ the 
HQET limit, replacing $\bar b$ field by an effective $\bar{h}_v$ field
and use the expansion in terms of $B$-meson DAs.
Their definitions, twist assignment and dependence on the momentum variable  are  
presented in Appendix~\ref{sect:appBDA}. Here the index $v$ indicates the velocity
four-vector $v=(P+q)/m_B$. We arrive at the following intermediate expression for the OPE of the correlation function: 
\begin{eqnarray}
&&{\cal F}^{(d)_{OPE}}(P,q)= 2\Bigg(\frac{i f_B m_{B}}{8}\Bigg)\int_{0}^{\infty}\!\!\! d\omega\,  \!\! \int \!d^4x\ e^{i(P-\omega v)\cdot x} \frac{x_{\alpha} x_{\beta}} {(2\pi^{2})^{2} (x^2)^4} 
\nonumber \\
&&\times \Big\{\!\! -\text{Tr}\! [\gamma^\mu \gamma^\alpha (1+\slashed{v})] \phi^B_+(\omega) + \text{Tr}[\gamma^{\mu} \gamma^\alpha (1+\slashed{v})\slashed{v} \slashed{x}] \frac{i}{2}\Phi^B_{\pm}(\omega) \Big\}\gamma_{5} \gamma_{\mu} \gamma_{\beta}(1+\slashed{v}) + \dots,
\label{eq:corrqq2}
\end{eqnarray}
 where for brevity we only show the contributions of the twist-2 and twist-3 DAs.
Note that a factor of 2 is inserted in the above to account for the two 
possible contractions of the $u$-quark fields in (\ref{eq:corr}),  leading to equal results.
Performing the Dirac-matrix algebra and integrating 
over the four-coordinate $x$,
we finally obtain:
\begin{align}
  \nonumber  {\cal F}^{(d)_{OPE}}(P,q)&=\frac{ f_B m_B}{16\pi^2}\!\int\limits_{0}^{\infty}\! d\omega 
  \Bigg[(P-\omega v)^2 \ln\!\big[\!-\!(P-\omega v)^2\big] \phi^B_+(\omega) \nonumber\\  
  & - \ln\!\big[\!-\!(P-\omega v)^2\big]\big(\slashed{P} \slashed{v} - \omega\big)
  \Phi^B_{\pm}(\omega) \nonumber\\ 
 & -8 \ln\big[\!-\!(P-\omega v)^2\big] g^B_+(\omega)  
+ 8\frac{\big(\slashed{P} \slashed{v} - \omega\big)}{(P-\omega v)^2}G_{\pm}^B(\omega)
  \Bigg] \gamma_L\,,  
  \label{eq:corrfinald}
\end{align}
where also the twist-4 and twist-5 DAs contribute.
Note that after the $x$-integration in (\ref{eq:corrqq2}) only the (convergent at $D=4$) terms generating imaginary part
over the variable $P^2$ are retained in (\ref{eq:corrfinald}), whereas the constant and polynomial terms in $P^2$ are omitted, since they will vanish
after Borel transform.

Following the same procedures as before, we calculate the correlation function in model $(b)$, and obtain a simple relation:
\begin{equation}
{\cal F}^{(b)_{OPE}}(P,q) = 2  {\cal F}^{(d)_{OPE}}(P,q).
\label{eq:formfactequal}
\end{equation}

Matching the result 
of our calculation in  (\ref{eq:corrfinald})
with the decomposition
(\ref{eq:corrdecomp}), we find that only the invariant amplitudes $F^{(d)}_{1}(P^2,q^2)$ and $F^{(d)}_{4}(P^2,q^2)$ contribute to
the adopted approximation of OPE.
We also  make the dependence on the momentum squared $P^2$ 
explicit, transforming:
\begin{equation} 
-(P-\omega v)^2=\Big(1-\frac{\omega}{m_B}\Big)\big( s(\omega) -P^2\big)\,,
\label{eq:Ptos}
\end{equation}
where a new variable 
\be
s(\omega)=\frac{\omega\big[m_B(m_B-\omega)-q^2\big]}{m_B-\omega}
\label{eq:somega}
\ee
is introduced,
so that $s(0)=0$ ,  $s(\infty)=\infty$ 
(with the asymptotic behavior $\lim_{\omega\to\infty} s(\omega) \sim m_B\omega$) 
and, inversely:
\be
\omega(s)=\frac{1}{2m_B}\Big(
m_B^2-q^2+s-\sqrt{(m_B^2-q^2+s)^2-4 m_B^2 s}\,
\Big)\,.
\label{eq:omegas}
\ee

Furthermore, since we are only interested in the 
terms in the expression
(\ref{eq:corrfinald}) that develop imaginary part in the variable $P^2$, we use (\ref{eq:Ptos})
and replace
$$\ln\!\big[\!-\!(P-\omega v)^2\big]\to \ln\big[ s(\omega)-P^2\big]\,.$$
The resulting
invariant amplitudes in the decomposition (\ref{eq:corrdecomp}) are:
\begin{align}
  \nonumber  {\cal F}^{(d)_{OPE}}_{1}(P^2,q^2)  
  &=\frac{ f_B}{16\pi^2}\!\int\limits_{0}^{\infty}\! d\omega 
  \Bigg[
  -(m_B-\omega)\big( s(\omega) -P^2\big)  
  \ln\!\big[ s(\omega)-P^2\big]\ \phi^B_+(\omega) 
  \nonumber\\  
  & - \ln\!\big[ s(\omega)-P^2\big]\big(P^2 - \omega m_B\big)
  \Phi^B_{\pm}(\omega) 
  \nonumber\\ 
 & -8 m_B \ln\big[ s(\omega)-P^2\big] g^B_+(\omega)  
-\frac{8\big(s(\omega)  - \omega m_B\big)}{
\Big(1-\frac{\omega}{m_B}\Big)\big( s(\omega) -P^2\big)
}\,G_{\pm}^B(\omega)
  \Bigg]\,,  
  \label{eq:corrF1}
\end{align}
where, in the coefficient multiplying $G^B_{\pm}$, the term that does not contribute to the imaginary part is removed, replacing $P^2\to s(\omega)$
in the numerator, and 
\begin{align}
 {\cal F}^{(d)_{OPE}}_{4}(P^2,q^2)  
  &=\frac{ f_B m_B^2}{16\pi^2}\!\int\limits_{0}^{\infty}\! d\omega 
  \Bigg[
   - \ln\!\big[ s(\omega)-P^2\big]
  \Phi^B_{\pm}(\omega) 
  -\frac{8G_{\pm}^B(\omega)}{
 \Big(1-\frac{\omega}{m_B}\Big)\big( s(\omega) -P^2\big)}
  \Bigg]\,.  
  \label{eq:corrF4}
\end{align}
All remaining invariant amplitudes vanish,
that is ${\cal F}^{(d)}_{2,3}(P^2,q^2)={\cal F}^{(d)}_{5,6,7,8,}(P^2,q^2)=0$
in the adopted LO and twist-5
approximation \footnote{Since the amplitudes 
${\cal F}^{(d)}_{5,6,7,8}$ belong to the Dirac structures with $\gamma_R$ 
in the decomposition (\ref{eq:corrdecomp}),
we expect that their absence  is not related to the accuracy of OPE, but is rather  correlated with the chirality properties of the underlying operator $\bar{O}_{(d)}$. This conjecture deserves further investigation which is beyond our scope here.}.

The next task is to transform 
(\ref{eq:corrF1}) and (\ref{eq:corrF4})
to the form of a dispersion integral (\ref{eq:OPEdsip}).
The details of that procedure are explained 
in Appendix~\ref{app:disp}.
Applying it, we calculate the imaginary parts of 
the invariant amplitudes  entering the 
integrals on right-hand sides of the LCSRs (\ref{eq:lcsr1}) and (\ref{eq:lcsr4}):
\begin{eqnarray}  
&&\text{Im} {\cal F}_1^{(d)_{OPE}}(s,q^2)
=\frac{ f_B}{16\pi}\! \Bigg\{\int\limits_{0}^{\omega(s)}\! d\omega' 
  \Bigg[\bigg( \omega'\big[m_B(m_B-\omega')-q^2\big] 
  -(m_B-\omega')s\bigg)  \phi^B_+(\omega') 
\nonumber\\  
&&  +\big(s - m_B\,\omega'\big)
  \Phi^B_{\pm}(\omega') 
+8m_B\, g^B_+(\omega')\Bigg]
-\frac{8m_B\big(s - m_B\,\omega(s)\big)}{
m_B-\omega(s)}
\, \frac{d\omega(s)}{ds} G_{\pm}^B(\omega(s))\,\Bigg\}
  \,,  
\label{eq:imF1}
\end{eqnarray}
\ba
\text{Im} {\cal F}_4^{(d)_{OPE}}(s,q^2)
=\frac{ f_B m_B^2}{16\pi}\Bigg\{\!\int\limits_{0}^{\omega(s)}\! d\omega' 
\Phi^B_{\pm}(\omega') -
\frac{8m_B\,G_{\pm}^B(\omega(s))}{
 m_B-\omega(s)}\frac{d\omega(s)}{ds}
 \Bigg\} \,,  
\label{eq:imF4}
\ea
where the $q^2$-dependence of the function $\omega(s)$ defined  in (\ref{eq:omegas}) is not shown for brevity. Note that in the adopted accuracy of OPE,
the remaining contributions to these sum rules vanish:
$$\text{Im} {\cal F}_2^{(d)_{OPE}}(s,q^2)=\text{Im} {\cal F}_3^{(d)_{OPE}}(s,q^2)=0\,.$$

\subsection{Contribution of three-particle DAs}
\label{sect:ope_3DAs}
\begin{figure}[t]
\begin{center}
\centering
\includegraphics[scale=0.5]{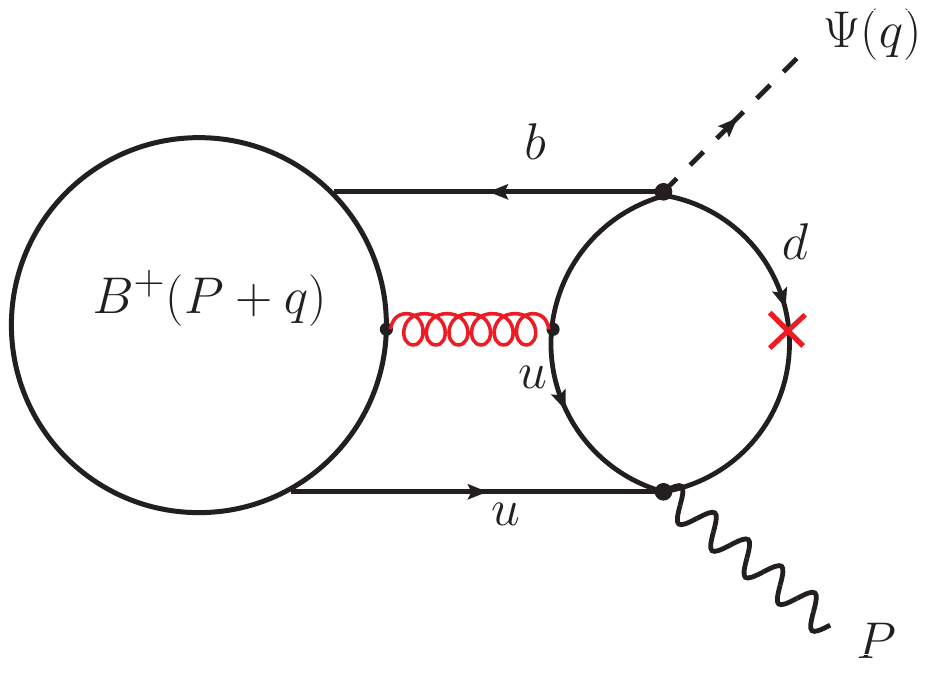}
\end{center}
\caption{Diagram  describing the contributions of three-particle (quark-antiquark-gluon) $B$-meson DAs 
to the correlation function (\ref{eq:corr}). The curly line represents a gluon component of the DA. The cross indicates the 
second diagram, where the gluon is attached to the  $d$-quark.}
\label{fig:diagG}
\end{figure}

The contributions of $B$-meson three-particle DAs to OPE are described by the diagrams shown in Fig.~\ref{fig:diagG}.
To compute  them, we need 
the quark propagator expanded  near the light-cone
 up to  the first-order  in gluon-field strength. In the adopted limit  of massless $u,d$ quarks, the corresponding expression reads
\cite{Balitsky:1987bk}:
\begin{equation}
\begin{aligned}
-i\langle 0| T\{q^i_\alpha(x)\,\bar{q}^k_\beta(0)\} |0\rangle
&= \frac{\slashed{x}_{\alpha\beta}\delta^{ik}}
       {2\pi^2(-x^2)^{2}}  \\
&\quad + \frac{g_s (\lambda^a)^{ik}}
       {32\pi^2(-x^2)}
       \int_0^1 du 
\Big\{
u\,\sigma_{\mu \nu} G^{a\mu\nu}(ux)\slashed{x}
+ \bar{u}\,\slashed{x}\sigma_{\mu \nu} G^{a\mu\nu}(ux)
\Big\}_{\alpha\beta} \, ,
\end{aligned}
\label{eq:prop}
\end{equation}
where $\lambda^a$ are the Gell-Mann matrices and 
the Dirac and color indices are shown explicitly.

To obtain the contribution of the first diagram, where the gluon from the $B$-meson DA is attached to  the virtual $u$-quark,
we use the second term in (\ref{eq:prop}), while using the free 
propagator for  the virtual $d$-quark. For the second diagram, these two terms are interchanged. As in the case of the LO diagram, 
the two possibilities to contract $u$-quark lines originating from the operator $\overline{\cal O}_{(d)}$ and interpolating current yield an overall factor of 2.
The three-particle DAs that arise when the quark, antiquark, and gluon fields are sandwiched between the $B$-meson and vacuum states are described in Appendix~\ref{sect:appBDA}.
Our result for the diagram with gluon attached to the $u$-quark
reads: 
\begin{equation}
\mathcal{F}^{(d)}_{\text{OPE}}(P,q)\big|_{u} = -\frac{f_B m_B}{64\pi^2} \int_0^\infty \!\! dw_1 \int_0^\infty \!\! dw_2 \int_0^1 \!\! \frac{du}{{\cal P}^2} \left[ 
u\, {\cal C}({\cal P},v) + \bar{u}\, \overline{{\cal C}} ({\cal P},v) \right] \gamma_L\,,
\label{eq:3partdiag}
\end{equation}
where  ${\cal P} \equiv P-
\sigma m_B v$, with the function $\sigma=\sigma (u,\omega_1,\omega_2)$ defined in (\ref{eq:sigma}). 
The lengthy expressions for the coefficients ${\cal C}$ and $\overline{{\cal C}}$
are presented in (\ref{eq:3partdiag1}) in  Appendix~\ref{sect:appBDA}. 

The calculation of the remaining diagrams
yields the following relations:
\begin{equation}
\begin{aligned}
&{\cal F}^{(d)_{OPE}}(P,q)|_{d}= -{\cal F}^{(d)_{OPE}}(P,q)|_{u}\,,
\end{aligned}
\label{eq:3partrel1}
\end{equation}
\begin{equation}
\begin{aligned}
&{\cal F}^{(b)_{OPE}}(P,q)|_{d}= {\cal F}^{(b)_{OPE}}(P,q)|_{u} = 0.
\end{aligned}
\label{eq:3partrel2}
\end{equation}
Thus, the contribution of 
three-particle DAs to the correlation function
vanishes for both models $(d)$ and $(b)$.
The reason, as follows from a detailed inspection of the diagram expressions, lies in the specific Dirac structure of the chosen proton interpolating current. Note that Ioffe current is a definite linear combination  of two other 
 three-quark currents (see e.g. \cite{Espriu:1983hu} for details). 
  We checked that  both  of them produce nonvanishing contributions which then 
  cancel  or vanish in this combination. 
  Furthermore, the Dirac structure of 
 the effective interactions given by the operators (\ref{eq:Od}) and 
 (\ref{eq:Ob}) coupled to $\Psi$ plays a role too. 
 Note that  in these  interactions both diquark and quark-$\Psi$ fields form Lorentz-scalars. 
 If the effective interaction would have another  Dirac structure, e.g., 
 $\left(\bar{u}^i_{R} \gamma_\rho b_R^{c}\right)\left(\bar{d}^{}_R\gamma^\rho \Psi\right)$ in model $(d)$,  (or the same with $d\leftrightarrow b$ in model $(b)$), the contributions of
 the diagrams  in Fig.~\ref{fig:diagG} would not 
 vanish in both models, 
 even in combination with the Ioffe current. 
  
  As a cross-check, we recalculated the 
  diagrams in Fig.~\ref{fig:diagG} with  quark propagators in the momentum representation and in $D\neq 4$. We found that, in  our case of 
  massless $u,d$ quarks, 
  the same relations (\ref{eq:3partrel1})
  and (\ref{eq:3partrel2}) are valid. On the other hand,
replacing in the correlation function the massless $d$-quark  by a massive $s$-quark  
  (which would correspond to 
  a decay mode $B\to\Lambda \Psi$), yields nonvanishing
  terms  proportional to $m_s$ for the contributions of three-particle DAs.

\section{Numerical results}
\label{sect:num}
\subsection{Inputs}
\begin{table}[ht]
\centering
\begin{tabular}{|l|l|c|}
\hline
Parameters  &  values/intervals & Source \\[1mm]
\hline 
\hline
&&\\[-3.0mm]
Hadron masses   & $m_B=5.279 $ GeV
 & \cite{ParticleDataGroup:2024cfk}
 \\[1mm]
& $m_p=938.272$  MeV &
\\[1mm]
& $m_{\Lambda_b}=5.620$  GeV &
\\[1mm]
& $m_{\pi^+}=139.57$  MeV &
\\[1mm]
 \hline
&&\\[-3.0mm]
Nucleon decay constant &  
$\lambda_p =-(27\pm 9)\times 10^{-3}$ 
GeV$^{\,2}$
 & \cite{Lenz:2009ar}\\[1mm]
\hline
$B$-meson decay constant & $f_B=190.0\pm 1.3$ MeV&\cite{FlavourLatticeAveragingGroupFLAG:2024oxs}
\\[1mm]
Inverse moment of tw2 $B$ DA  & 
 $\lambda_B =460 \pm 110$ MeV  
 &\cite{Braun:2003wx}\\[1mm]
 The ratio (\ref{eq:R}) 
& $R=\lambda_E^2/\lambda_H^2=0.30^{+0.17}_{-0.12}$ & see App.~\ref{sect:appBDA}\\[1mm]
\hline
&&\\[-3.0mm]
Borel parameter squared & $M^2= 1.5$ - $2.0 $ GeV$^2$ &
\cite{Anikin:2013aka}
\\[1mm]
Duality threshold  & $ s_0 = 2.25 $ GeV$^2$ &  \\[1mm]
\hline
&&\\[-3.0mm]
$b$-quark $\overline{\text{MS}}$ mass & $\overline{m}_b(\mu=3~\mbox{GeV})= 4.47^{+0.04}_{-0.03} $ GeV  & 
\cite{ParticleDataGroup:2024cfk}  \\[1mm]
\hline
&\\[-3.0mm]
$B^+$-meson lifetime & $\tau_B= 1.638\pm 0.004$ ps  & 
\cite{ParticleDataGroup:2024cfk}  \\[1mm]
\hline
\hline
\end{tabular}
\caption{The input parameters used in the numerical analysis.  }
\label{tab:input}
\end{table}

In Table~\ref{tab:input}, we collect the values and intervals of all input parameters used in  our numerical analysis. For 
LCSRs, apart from hadron masses taken from \cite{ParticleDataGroup:2024cfk}, 
we need the proton decay constant 
$\lambda_p$  defined in 
(\ref{eq:deconst}). For that we take the interval 
from \cite{Lenz:2009ar},
determined from two-point QCD sum rule for the same Ioffe currents.

Definitions and models adopted for the $B$-meson DAs entering LCSRs are described in Appendix~\ref{sect:appBDA}. 
The whole set of these DAs contains three input parameters, which  currently have 
different levels of precision. The $B$-meson decay constant is very 
accurately determined from lattice QCD calculations.
We adopt the  average of $N_f=3+1+1$  results from  \cite{FlavourLatticeAveragingGroupFLAG:2024oxs}.
Still far more uncertain is the inverse moment of 
the twist-2 $B$-meson DA defined in (\ref{eq:lamB}). We use the 
determination of this parameter  from QCD sum rule based on local OPE \cite{Braun:2003wx} (see also \cite{Khodjamirian:2020hob}). 
The third input for $B$-meson DAs is the ratio (\ref{eq:R}),
for which the interval in Table~\ref{tab:input} is obtained by 
averaging the parameters $\lambda_{E}$ and $\lambda_H$ 
over three independent determinations \cite{Grozin:1996pq}, 
\cite{Nishikawa:2011qk}, \cite{Rahimi:2020zzo} from 
QCD sum rules based on  different two-point correlation functions. Adopting the values quoted 
in Table~\ref{tab:input}, we tacitly choose for $\lambda_B$ and other scale-dependent parameters
one and the same renormalization scale $\mu\sim 1$ GeV,
 which lies in the ballpark of the Borel mass scale.
 More accurate choice
of scales demands computation of gluon radiative corrections to the 
correlation function, which is a technically quite a demanding task remaining beyond our scope. 
The choice of Borel-mass parameter and of effective threshold
presented in Table~\ref{tab:input} is discussed in the next subsection. Finally, for the estimate of the inclusive width 
$B\to X_N \Psi$ used below in Section~\ref{sect:excl-incl},
we use the $b$-quark mass from \cite{ParticleDataGroup:2024cfk} at the scale $\mu = 3$ GeV.

\subsection{Numerical analysis of LCSRs }

According to our results for the OPE obtained in Section~\ref{sect:ope}, 
within the adopted approximation only the two invariant amplitudes 
${\cal F}_{1}^{(d)}$ and ${\cal F}_{4}^{(d)}$ 
contribute to the LCSRs derived in Section~\ref{sect:lcsr}, 
whereas all other amplitudes vanish.  Hence, here we concentrate on the numerical analysis of the two 
sum rules (\ref{eq:lcsr1}) and (\ref{eq:lcsr4}).

Furthermore, as already discussed above, we 
simplify the hadronic parts of these sum rules, 
including all nucleon resonance contributions into the part of the 
hadronic spectral density  that is approximated with the quark-hadron duality, applying (\ref{eq:dual}). In this respect, 
we follow the analysis of the  
LCSRs for the nucleon electromagnetic form factors
(see e.g.~\cite{Anikin:2013aka,Braun:2001tj}). It is then natural to adopt the same effective threshold 
$s_{0(N)}\simeq (1.5~\mbox{GeV})^2=2.25~\mbox{GeV}^2$ and the same interval of Borel mass squared, $M^2=1.5-2.0$ GeV$^2$.

From the LCSRs (\ref{eq:lcsr1})
and (\ref{eq:lcsr4})
we finally obtain the $B\to p$ form factors 
\be
F^{(d)}_{B\to p_R}(q^2)
=\frac{1}{\pi\lambda_p m_p^2}\int\limits_{0} ^{s_{0(N)}}\!\! ds\,e^{\frac{m_p^2-s}{M^2}}
\,\text{Im} {\cal F}_1^{(d)_{OPE}}(s,q^2)\,, 
\label{eq:lcsrfin1}
\ee
\be
\widetilde{F}^{(d)}_{B\to p_L}(q^2)
=\frac{1}{\pi\lambda_p m_B^2}\int\limits_{0} ^{s_{0(N)}}\!\! ds\,e^{\frac{m_p^2-s}{M^2}}
\,\text{Im} {\cal F}_4^{(d)_{OPE}}(s,q^2)\,. 
\label{eq:lcsrfin4}
\ee
We parenthetically note that the LCSRs 
(\ref{eq:lcsr2}) and (\ref{eq:lcsr3})
with a vanishing OPE part
indicate 
nontrivial cancellation 
between  the $B\to p$, $B\to N^*(1535)$
and $B\to N(1440)$ form factors, or in other words, relations between these form factors which deserve a separate
study.
For the form factors in model $(b)$, we have the same 
LCSRs (\ref{eq:lcsrfin1}) and (\ref{eq:lcsrfin4}),
where the right-hand side is multiplied by a factor 2,
as implied by the relation (\ref{eq:formfactequal}). Hence:
\be
F^{(b)}_{B\to p_R}(q^2) = 2F^{(d)}_{B\to p_R}(q^2)\,,~~
\widetilde{F}^{(b)}_{B\to p_L}(q^2)=
2\widetilde{F}^{(d)}_{B\to p_L}(q^2)\,.
\label{eq:rel}
\ee

\begin{figure}[htbp]
    \centering
    \includegraphics[width=0.475\textwidth]{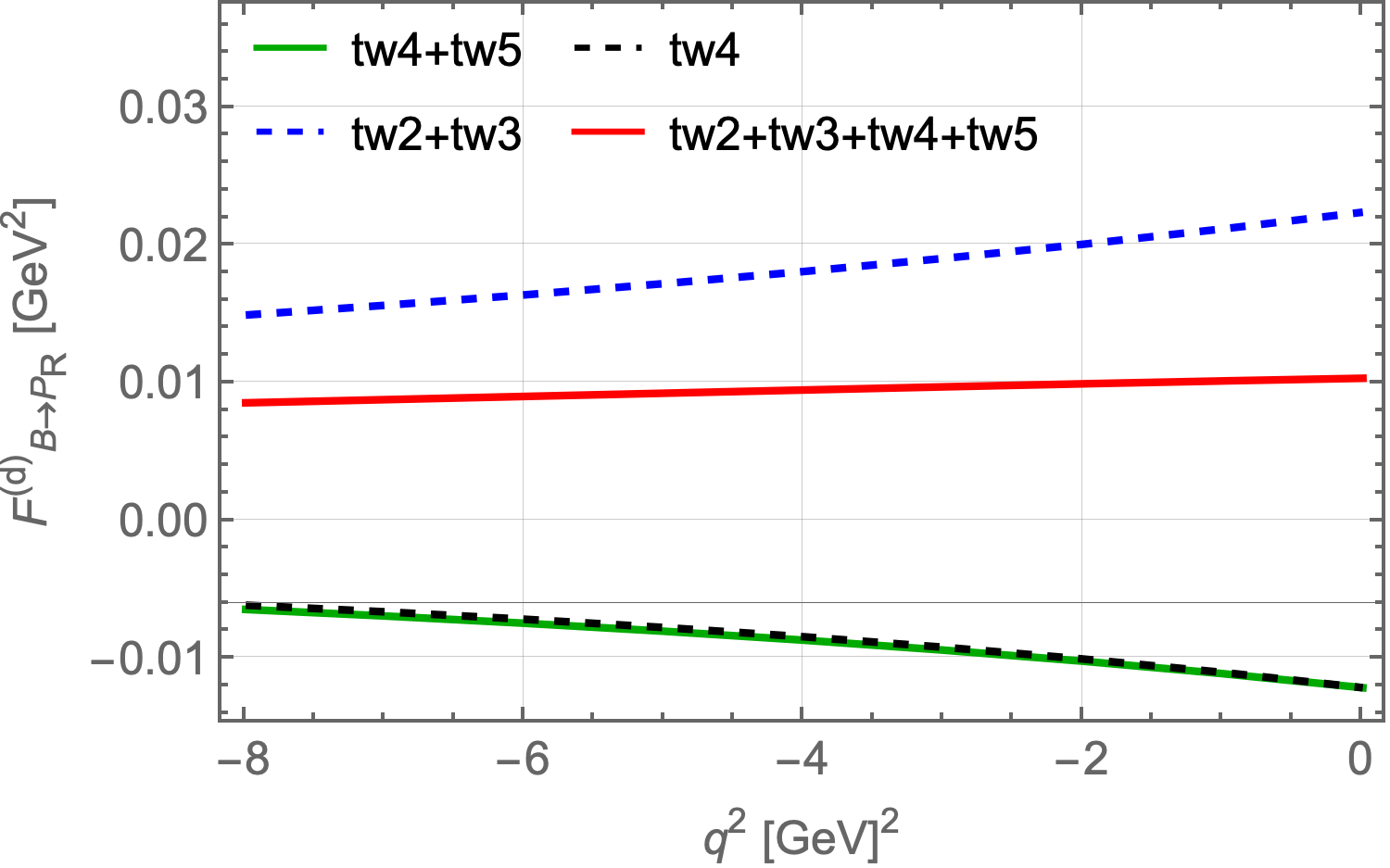}
    \hfill
    \includegraphics[width=0.49\textwidth]{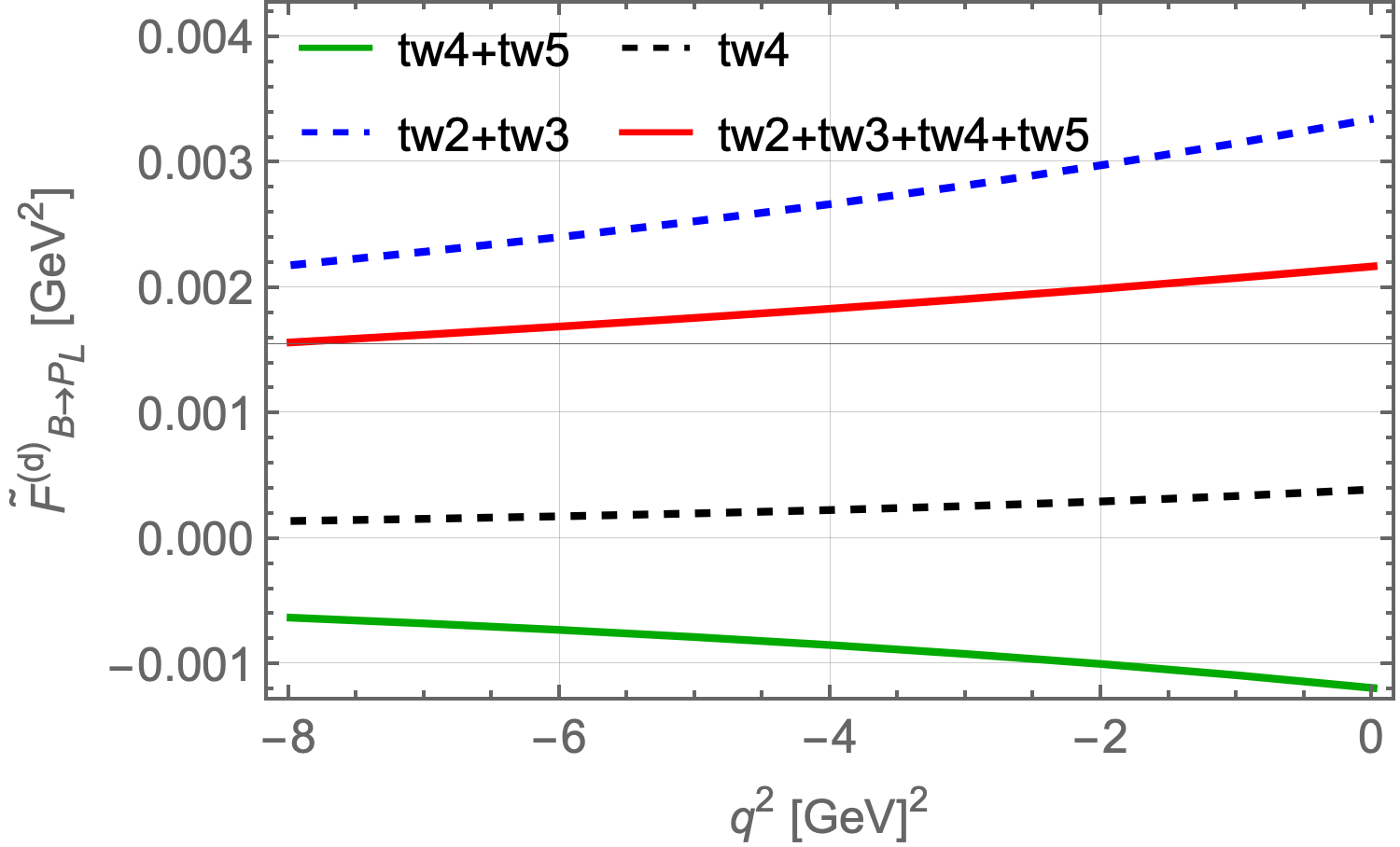}
\caption{The form factors  $F^{(d)}_{B\to P_R}$ (left panel) and $\widetilde{F}^{(d)}_{B\to P_L}$ (right panel) in the spacelike region and at central input,   shown  with the red solid lines.
 The blue dashed (green solid)
lines represent 
the sum of the twist-2 and twist-3 (twist-4 and twist-5) contributions, and the black dashed line represents the twist-4 
contribution.
}
\label{fig:DA_compar}
\end{figure}

\begin{figure}[h]
    \centering
    \includegraphics[width=0.475\textwidth]{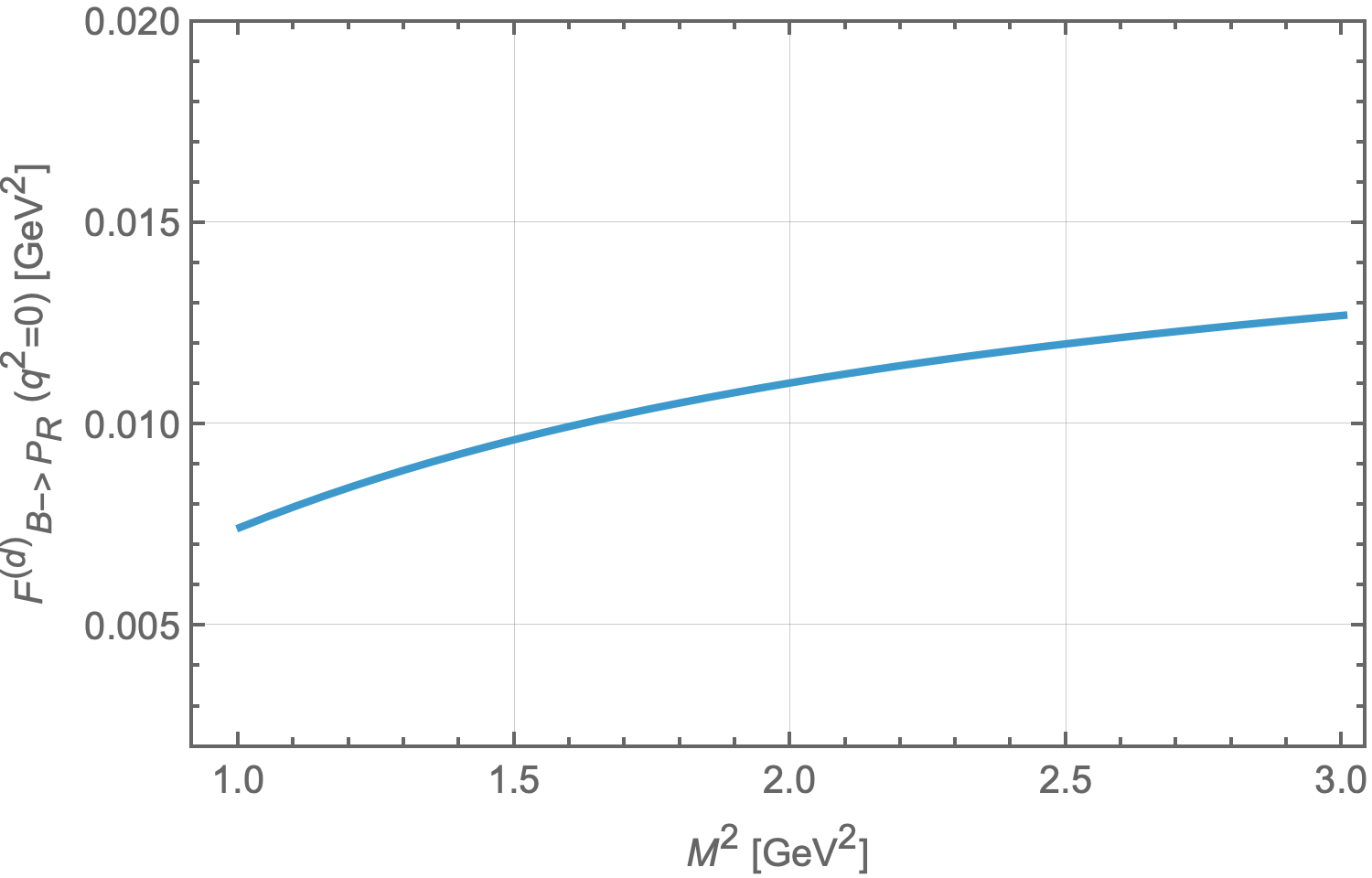}
    \hfill
    \includegraphics[width=0.475\textwidth]{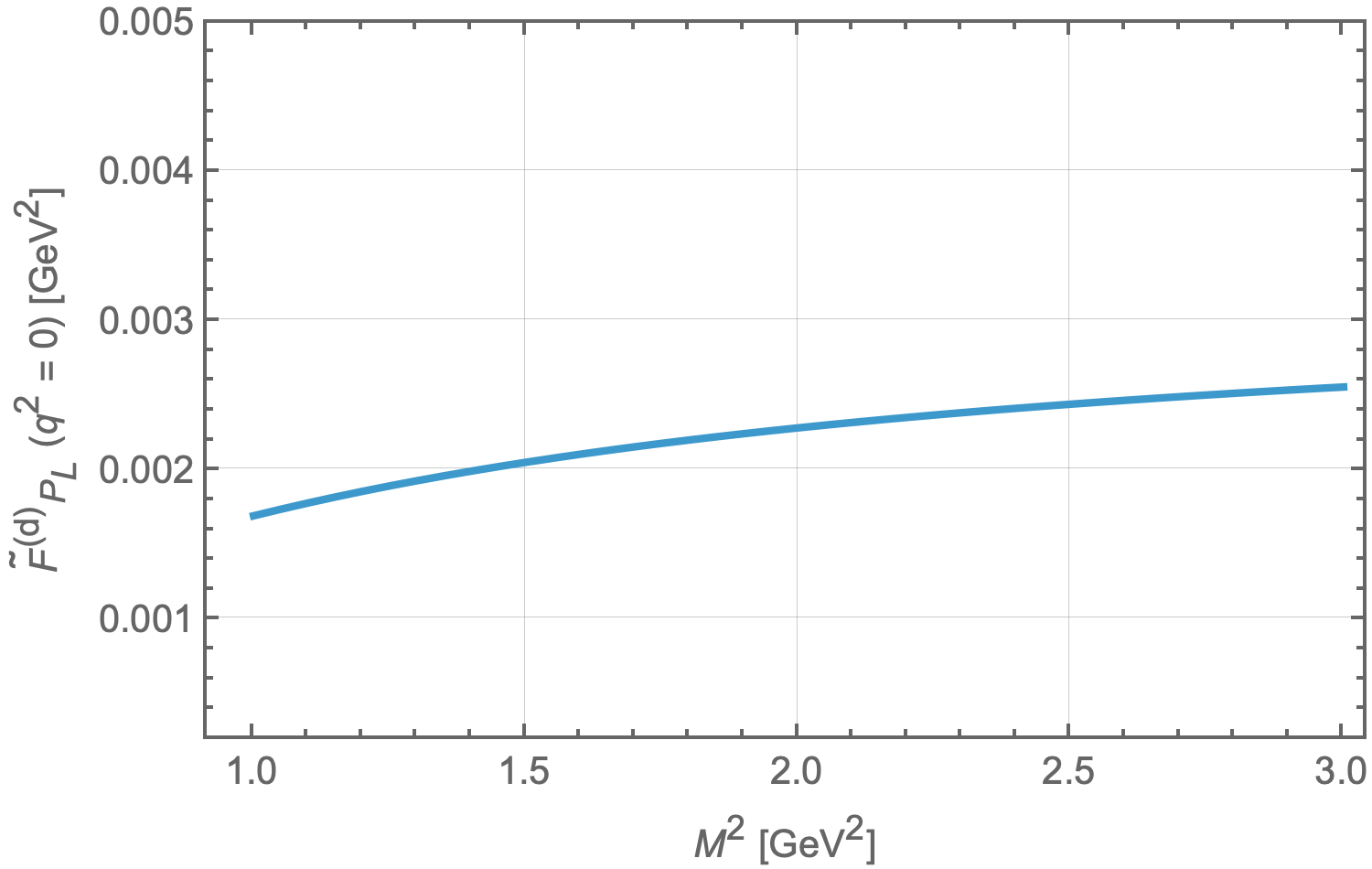}
\caption{Borel-mass dependence of the form factors $F^{(d)}_{B\to P_R}$ (left panel) and 
$\widetilde{F}^{(d)}_{B\to P_L}$ (right panel) at $q^2=0$, evaluated at the central values of the input parameters.
The interval $M^2=1.5 -2.0~\mbox{GeV}^2$ is used in LCSRs. 
}
\label{fig:FF_Borel}
\end{figure}

Having at hand all input parameters, we evaluate the form factors
given by the LCSRs (\ref{eq:lcsrfin1})  and  (\ref{eq:lcsrfin4})  in the region of their validity, $q^2<0$. The numerical values of both form factors $F^{(d)}_{B\to p_R}(q^2)$ and 
$\widetilde{F}^{(d)}_{B\to p_L}(q^2)$ at the central input  
and in the  interval $-8.0~\mbox{GeV}^2\le q^2\le 0$\,
are plotted  in Fig.~\ref{fig:DA_compar}.
To illustrate the convergence of OPE, we also display in the same figure
the contributions of $B$-meson DAs with separate twists.
Since the twist-3 DA $\phi^B_- $ enters LCSRs only 
in  the integrated linear combination  with 
the twist-2 DA  $\phi^B_+$, we add together their
contributions. For the same reason, we plot the sum
of twist-4 and twist-5 contributions. We also  separately display  the  contribution of twist-4 DA, to assess the influence of the twist-5 term on LCSRs. 
Furthermore, we show the dependence of the form factors on the Borel parameter $M^2$ in Fig.~\ref{fig:FF_Borel},
indicating a reasonably stable behaviour also beyond  
the interval chosen for LCSRs. 

Comparing separate contributions to LCSRs, we find that for the form factor $F^{(d)}_{B\to P_R}$, across the entire $q^2$ interval,
the DAs of higher twists 4,5  have smaller contributions than the DAs of lower twists 2,3,
while the effect of adding the twist-5 DA is negligible. A similar hierarchy is observed for the form factor $\widetilde{F}^{(d)}_{B\to P_L}$, however, here the 
twist-5 DA has  a larger contribution than its twist-4 counterpart. For model $(b)$, the identical pattern appears scaled by a factor of two, as follows from the relation (\ref{eq:rel}). 
In addition, our numerical analysis reveals,  
that applying a simplified version of hadronic part
with the effective  threshold $s_0$ covering  nucleon
resonances affects the  proportion 
between  the proton-pole term and the integral above $s_0$
in the LCSRs. The pole  term contribution to the sum rule 
becomes subdominant, which means that our LCSR results 
have a substantial reliance on the quark-hadron duality approximation. 

Up to this point, the numerical evaluation of the 
form factors from LCSRs was done at the central values of the input parameters. We also estimate total uncertainties of our predictions arising from the variation of input parameters within adopted intervals.
The uncertainties of the form factors are obtained by adding the individual parametric variations in quadrature, and by propagating the resulting upper and lower variations.

In Table~\ref{tab:compar}, we present, as a sample of our results,
 the $B \to p$ transition form factors at $q^2 = 0$. 
The final results (denoted as LCSR-$B$)  are in the column
(tw2-tw5), while the column (tw2-tw3) is just for illustration. The resulting uncertainties are quite large, mainly due to the 
broad intervals of the two key input parameters:  
the inverse moment $\lambda_B$ of the leading twist-2 DA, 
and the nucleon decay constant $\lambda_p$. Note that we prefer to display our predictions in this Table in the form of 
intervals, because for an LCSR 
numerical result no Gaussian distribution can be attributed to the 
most important input uncertainties. Indeed, both $\lambda_B$ and $\lambda_p$ 
are themselves derived from two-point QCD sum rules. 

\begin{table}[t]
\centering
\begin{tabular}{|c|c|c||c|c|}
\hline
\multirow{2}{*}{Form factor} & \multicolumn{2}{c||}{LCSR-$B$ (this work)} & \multicolumn{2}{c|}{LCSR-$N$} \\
\cline{2-5}
& (tw2-tw3) & (tw2-tw5) & (tw3) \cite{Khodjamirian:2022vta} & (tw3-tw6) \cite{Boushmelev:2023huu} \\
\hline
\hline
$F^{(d)}_{B\to p_R}(0)$ 
& $[1.4,\,3.9]$ & $[0.4,\,2.2]$ 
& $[2.4,\,2.8]$ & $[0.9,\,3.5]$ \\
\hline
$\widetilde{F}^{(d)}_{B\to p_L}(0)$ 
& $[0.24,\,0.52]$ & $[0.14,\,0.35]$ 
& $0$ & $[0.4,\,0.7]$ \\
\hline
\hline
$F^{(b)}_{B\to p_R}(0)$ 
& $[2.8,\,7.7]$ & $[0.8,\,4.4]$ 
& $[-0.9,\,-0.6]$ & $[-5.9,\,-2.2]$ \\
\hline
$\widetilde{F}^{(b)}_{B\to p_L}(0)$ 
& $[0.5,\,1.1]$ & $[0.28,\,0.70]$  
& $[0.07,\,0.10]$ & $[-0.9,\,-0.4]$ \\
\hline
\end{tabular}
\caption{Intervals of the $B \to p$ transition form factors at $q^2 = 0$ (in units of $10^{-2}$ GeV$^2$) for models $(d)$ and $(b)$, obtained from LCSRs with $B$-meson DAs (denoted as LCSR-$B$), are compared with the results from LCSRs with 
nucleon DAs (LCSR-$N$). Here the notation such as (tw$2$--tw$5$) indicates that contributions from twist-$2$ up to twist-$5$ are included.}
\label{tab:compar}
\end{table}

\begin{table}[t]
\centering
\begin{tabular}{|c|c|c||c|c|}
\hline
\multirow{2}{*}{Form factor} & \multicolumn{2}{c||}{Uncertainty from $\lambda_B$} & \multicolumn{2}{c|}{Uncertainty from $\lambda_p$} \\
\cline{2-5}
& (tw2-tw3) & (tw2-tw5) & {(tw2-tw3)} & (tw2-tw5)  \\
\hline
\hline
$F^{(d)}_{B\to p_R}(0)$ 
& {$[1.6,\,3.5]$} & {$[0.5,\,2.0]$} 
& {$[1.7,\,3.4]$} & {$[0.8,\,1.6]$} \\
\hline
{$\widetilde{F}^{(d)}_{B\to p_L}(0)$} 
& {$[0.29,\,0.39]$} & {$[0.16,\,0.29]$} 
& {$[0.25,\,0.49]$} & {$[0.16,\,0.33]$} \\
\hline
\hline
{$F^{(b)}_{B\to p_R}(0)$} 
& {$[3.2,\,6.9]$} & {$[1.0,\,4.1]$} 
& {$[3.3,\,6.7]$} & {$[1.6,\,3.1]$} \\
\hline
{$\widetilde{F}^{(b)}_{B\to p_L}(0)$} 
& {$[0.59,\,0.77]$} & {$[0.32,\,0.59]$}  
& {$[0.5,\,1.0]$} & {$[0.33,\,0.65]$} \\
\hline
\end{tabular}
\caption{The same form factors from LCSRs with $B$-meson DAs
as in Table~\ref{tab:compar}. Shown are the intervals stemming from  separate variations of the parameter $\lambda_B$ or $\lambda_p$ with all other input parameters fixed at their central values.}
\label{tab:lambdaBpdev}
\end{table}

In addition, we estimated the variations of the 
form factors $F^{(d,b)}_{B\to p_R}$ and
$\widetilde{F}^{(d,b)}_{B\to p_L}$  
 at $q^2=0$, generated by the uncertainties of the parameters $\lambda_{B}$ and $\lambda_p$, varying each  of the latter separately and fixing all the other input parameters to their central values.
The resulting  intervals for the form factors are given in Table~\ref{tab:lambdaBpdev}.
Both $\lambda_B$-induced  and $\lambda_p$-induced uncertainties indeed constitute a large part of the total intervals quoted in Table~\ref{tab:compar}.  At the same time, both form factors in (\ref{eq:lcsrfin1}) and (\ref{eq:lcsrfin4})
depend on the proton decay constant through the same overall normalization factor 
$1/\lambda_p$. Hence, their ratio is independent of this parameter. 

Furthermore, we observe that including higher-twist contributions reduces the overall uncertainty. This originates from partial cancellations between leading- and higher-twist terms, which decrease both the central values of the form factors and their sensitivity to input parameters (see Fig.~\ref{fig:DA_compar}).

In Table~\ref{tab:compar}, we also quote previous results for 
the $B\to p$ form factors obtained from LCSRs employing 
the nucleon DAs, denoted as LCSR-$N$. For the latter sum rules, the initial  twist-3 
accuracy  in \cite{Khodjamirian:2022vta} was upgraded in \cite{Boushmelev:2023huu}  up to the twist-6  
level. Comparing both methods and their results with each other, we first of all notice that 
the simple relations (\ref{eq:rel}) between the form factors in models $(d)$ and $(b)$ predicted above are absent in the LCSR-$N$ 
at any twist. Moreover, in the latter approach both $(b)$-model
form factors have a different sign with respect to the ones 
in model $(d)$.

We emphasize that the concept of twist 
for the nucleon DAs is completely different from the one used 
 for the $B$-mesons DAs. Therefore, a direct one-to-one comparison of separate twist
contributions obtained with both sum rule methods is pointless as such.
Still, an overall suppression of  larger-twist contributions
remains a general criterion of validity, which is also satisfied by our  LCSR-$B$ results.  On the contrary, according  to \cite{Boushmelev:2023huu},  the LCSR-$N$ method yields  large,  and even dominant contributions of higher twists. To test that in more detail,
 we used the sum rule expressions from \cite{Boushmelev:2023huu} in the spacelike region, together with the input adopted there, and compared  numerical contributions from nucleon DAs of different twist. Specifically, concentrating on model $(b)$, and taking $q^2=-4.0~\mathrm{GeV}^2$, where the twist expansion is expected to be valid, we found 
 for the form factor $F^{(b)}_{B\to p_R}(q^2)$, that the twist-4 (twist-5) contribution is four times larger (amounts to $60\%$) than the twist-3 one. For the second form factor  
 $\widetilde{F}^{(b)}_{B\to p_L}(q^2)$  in the same model $(b)$
 the ratio of twist-4 to twist-3 contributions reaches an order
 of magnitude. This significant enhancement explains  
 a rather large difference between the LCSR-$N$ form factors in the first and
 second column of Table ~\ref{tab:compar}, especially for model $(b)$. For model $(d)$ the effects of higher twists are much less pronounced, and the agreement
 with our estimates is also better.
 
 Furthermore, we also verified the 
resulting uncertainties of the form factors obtained  in 
 \cite{Boushmelev:2023huu} and quoted as intervals in Table~\ref{tab:compar}. To this end, 
we performed an independent Monte-Carlo study of these uncertainties and found agreement  at  $q^2=0$ with the intervals 
 in the last column of 
Table~\ref{tab:compar}. The fact that the latter intervals 
are also  broad can be traced
to a combination of enhanced  twist-4,5 effects in the LCSR-$N$ method with still uncertain parameters of the corresponding nucleon DAs.

Note that  the estimated large uncertainties of both LCSR 
methods yield  an effective  marginal  agreement between 
their predictions, in the form of an overlap between the intervals
of predicted form factor values. 
But the status of OPE, expressed via hierarchy of  twist contributions, remains questionable for the LCSR-$N$ approach, at least for model $(b)$, and deserves further investigation.

Our LCSR-$B$ results will be used in the next subsection for extrapolation
of the form factors to the physical region of 
the $B^+\to p \Psi$ decay.

\subsection{ Branching fraction of the $B^+\to p\Psi$ decay} 
\label{sect:BR}
The mass of the dark antibaryon $\Psi$ in the $B^+\to p \Psi$ decay can vary within  the interval  
\be
m_p\simeq 0.94~ \mbox{GeV} <m_\Psi<(m_B-m_p)\simeq 4.21 ~\mbox{GeV}\,, 
\label{eq:region}
\ee
 where  the lower limit follows from the constraint that is put  in mesogenesis models 
\cite{Elor:2018twp,Alonso-Alvarez:2021qfd}
to avoid the proton decay into $\Psi$.
Hence, in order to estimate the width of $B^+\to p \Psi$, we need to extrapolate the $B\to p$ form factors obtained from LCSRs at $q^2<0$ to the timelike region of $q^2=m_\Psi^2$
between $m_p^2$ and $(m_B-m_p)^2$.

The invariant form factors determining the $B\to p$  hadronic matrix element  in (\ref{eq:ff}) have the same analytic  properties as any other
hadron form factor. The variable $q^2$ can be continued from the region $q^2<0$  to the whole complex $q^2$-plane.  
Singularities of any $B\to p$ form factor  are located on the positive real axis and are
generated by intermediate hadronic states with the quantum numbers of $b$-baryons. 
The lowest singularity is the pole 
at $q^2=m_{\Lambda_b}^2$ associated with  the $\Lambda_b$-baryon. 
 The next isospin-zero state is a hadronic continuum 
 state    composed from $\Lambda_b $ and two pions. This state develops a branch point  at $q^2=(m_{\Lambda_b}+2m_\pi)^2$. Without going into details on other heavier states, we only
 notice that the threshold corresponding to 
 the  state of $B$-meson and proton at $q^2=(m_B+m_p)^2$ is 
 by $\simeq 300~\mbox{MeV}$ heavier than the state $\Lambda_b 2\pi$.
 
Taking all these comments into account, we apply to the $B\to p$ form factors a standard $z$-expansion method based on analyticity, which  is widely used for heavy-to-light form factors and was also employed in~\cite{Khodjamirian:2022vta} for the same $B\to p$ form factors.
For definiteness, we consider the form factor 
$F^{(d)}_{B\to p_R}(q^2)$, all others are treated in the same way.
We employ the transformation of  the variable $q^2$ to  a new variable z, defined as 
\be
z(q^2) = \frac{\sqrt{t_{+}-q^2}-\sqrt{t_{+}}}{
\sqrt{t_{+}-q^2}+\sqrt{t_{+}}}\,, 
\label{eq:zpar}
\ee
with $t_{+} = (m_{\Lambda_b}+2m_{\pi})^2$.
Note that, for simplicity, we put to zero the other (arbitrary) parameter $t_0$ of this transformation, hence the point  $q^2=0$  transforms to $z=0$.
The spacelike interval, in which the form factor $F^{(d)}_{B\to p_R}(q^2)$ and the other form factors are evaluated using LCSRs, and the physical $q^2$ region relevant for the $B^+\to p\Psi$ decay are mapped, respectively, as follows:
\ba
-8.0~ \text{GeV}^2 < q^2 < 0 &\to&  
0.083 \geq z \geq 0 \,,
\label{eq:q2toz1}\\
m_p^2<q^2<(m_B-m_p)^2&\to& -0.067 > z>-0.20\,.
\label{eq:q2toz2}
\ea
Since the lowest possible threshold 
\footnote{ Note that in \cite{Khodjamirian:2022vta,
Boushmelev:2023huu}, while  using
a similar $z$-expansion, the 
$B p$ threshold was tacitly assumed to be the lowest one, choosing $t_+=(m_B+m_p)^2$. Here we correct this oversight. At the same time  we checked that if we would  adopt the same higher threshold, the numerical results for the extrapolated form factors shift within 1\% from our results 
obtained with a correct choice.} 
is chosen for the parameter $t_+$, only 
the $\Lambda_b$-pole of the form factor remains within the whole region  $|z|<1$ on which the $q^2$-plane 
maps. To remove this pole, we multiply the form factors by  $(1-q^2/m_{\Lambda_b}^2)$.  
The resulting function of $z$, 
$(1-q^2(z)/m_{\Lambda_b}^2)F^{(d)}_{B\to p_R}(q^2(z))$
is then singularity-free 
at $|z|<1$. Owing to the fact that 
both intervals of $z$ variable
(\ref{eq:q2toz1}) and (\ref{eq:q2toz2})
are close to $z=0$,
we expand this function in the powers  of $z$ 
around  $z=0$ and approximate it by first few terms.

More specifically, we use the standard BCL form of the $z$-expansion~\cite{Bourrely:2008za}, retaining terms up to $O(z^2)$. 
We employ a convenient form of this parametrization  
(see e.g.,~\cite{Khodjamirian:2022vta}):
\begin{eqnarray}
F^{(d)}_{B\to p_R}(q^2)= \frac{F^{(d)}_{B\to p_R}(0)}{1 - q^2/m_{\Lambda_b}^2}
\Bigg\{1 + \beta^{(d)}_{B\to p_R}\Bigg[z(q^2) - z(0)
+ \frac{1}{2} \Big( z(q^2)^2 - z(0)^2\Big)\Bigg] \Bigg\}\,.
\label{eq:BCL}
\end{eqnarray}
in terms of two parameters: 
the form-factor value at $q^2=0$ and the slope parameter $\beta^{(d)}_{B\to p_R}$. A similar  expression with the corresponding slope parameter
$\tilde{\beta }^{(d)}_{B\to p_L}$ is used for the form factor
$\widetilde{F}^{(d)}_{B\to p_L}(q^2)$. For the model-$(b)$ form factors, the relations (\ref{eq:rel}) are again used, yielding 
\be
F^{(b)}_{B\to p_R}(0)=2F^{(d)}_{B\to p_R}(0),~~~~
\widetilde{F}^{(b)}_{B\to p_L}(0)=
2\widetilde{F}^{(d)}_{B\to p_L}(0)\,,
\label{eq:bF0}
\ee
and equalities of slope parameters:
\be
\beta^{(b)}_{B\to p_R}= \beta^{(d)}_{B\to p_R}, ~~~~
\widetilde{\beta }^{(b)}_{B\to p_L} =\widetilde{\beta}^{(d)}_{B\to p_L}.
\label{eq:bbeta0}
\ee
The next step is a fit of  the parameterization (\ref{eq:BCL}) to 
our LCSR results 
for the  form factor  $F^{(d)}_{B\to p_R}(q^2)$ 
at $q^2\leq 0$.
To this end, we perform two different procedures. First, we 
take the LCSR form factor at $q^{2}=0$ as a fixed 
external input and fit only
the single slope parameter
$\beta^{(d)}_{B\to p_R}$, using 
the form factor calculated at eight equidistant points  
in the interval (\ref{eq:q2toz1}).
The fit is repeated, varying 
the form factor within  the estimated 
parametric uncertainty induced by the LCSR input.
This variation is then propagated to determine the  upper and lower limits of the slope parameter
$\beta^{(d)}_{B\to p_R}$ in a form of an asymmetric error. We repeat the same fit procedure for the second form factor $\tilde{F}^{(d)}_{B\to p_L}(q^2)$ .
The resulting parameters of $z$-expansion 
are presented in Table~\ref{tab:FitParameters},
where for the $(b)$-model form factors we use
the relations (\ref{eq:bF0}) and (\ref{eq:bbeta0}).

This simple approach with a fixed 
normalization and fitted slope parameter of the form factor reflects the full range of the form factor uncertainties, however
the correlations between the two parameters 
of our $z$-expansion  are not explicitly included in the error propagation. To assess the impact of such correlations, we have also performed an alternative fit, in which the value of $F^{(d)}_{B\to p_R}(0)$ is treated as a nuisance parameter and varied within its uncertainty. In that case, the covariance matrix obtained from the Hessian automatically accounts for correlations between the normalization and the slope. We find that the resulting branching fractions agree within uncertainties with those obtained in the fixed-normalization approach, indicating that correlation effects are inessential.

\begin{table}[h]
\centering
\renewcommand{\arraystretch}{1.8}
\begin{tabular}{|c|c|c|c|}
\hline
$F^{(d)}_{B\to p_R}(0)$ & $1.0^{+1.2}_{-0.6}$ & $\beta^{(d)}_{B\to p_R}$ & $0.99^{+3.4}_{-1.4}$ \\
\hline
$\widetilde{F}^{(d)}_{B\to p_L}(0)$ & $0.22^{+0.13}_{-0.08}$ & $\widetilde{\beta}^{(d)}_{B\to p_L}$ & $-1.78^{+0.76}_{-0.17}$ \\
\hline
\hline\hline
$F^{(b)}_{B\to p_R}(0)$ & $2.0^{+2.4}_{-1.2}$ & $\beta^{(b)}_{B\to p_R}$ & $0.99^{+3.4}_{-1.4}$ \\
\hline
$\widetilde{F}^{(b)}_{B\to p_L}(0)$ & $0.44^{+0.26}_{-0.16}$ & $\widetilde{\beta}^{(b)}_{B\to p_L}$ & $-1.78^{+0.76}_{-0.17}$ \\
\hline
\end{tabular}
\caption{Parameters of the $z$-expansion (\ref{eq:BCL}) obtained from 
the fit of LCSRs. Presented are the
$B\to p$ form factors at $q^2=0$ (in units of $10^{-2}~ \mbox{GeV}^2$) and the dimensionless slope parameters.}
\label{tab:FitParameters}
\end{table}

Having determined  the form factors in the timelike region of $q^2=m_\Psi^2$ via the $z$-expansion, 
we can now predict the branching fraction for $B^+ \to p \Psi$ as a function of the $\Psi$ mass. The expression for the decay branching fraction in terms of the form factors  was already given in~\cite{Khodjamirian:2022vta}, e.g., 
 we have for model~($d$):
\begin{align}
\mathrm{BR}_{(d)}(B^+\to p\Psi)
&= |G_{(d)}|^2 
\Bigg\{
\Bigg[
\Big(F^{(d)}_{B\to p_R}(m_\Psi^2)\Big)^2
+ \frac{m_\Psi^2}{m_p^2}\Big(\widetilde{F}^{(d)}_{B\to p_L}(m_\Psi^2)\Big)^2
\Bigg]
\big(m_B^2-m_p^2-m_\Psi^2\big)
\nonumber \\
&\quad + 2 m_\Psi^2\,
F^{(d)}_{B\to p_R}(m_\Psi^2)\,
\widetilde{F}^{(d)}_{B\to p_L}(m_\Psi^2)
\Bigg\}
\,\frac{\lambda^{1/2}(m_B^2,m_p^2,m_\Psi^2)}{16\pi m_B^3} \tau_{B^+},
\label{eq:widthb}
\end{align}
where $\lambda(a,b,c)=a^2+b^2+c^2-2ab-2ac-2bc$ is the K\"allen function and $\tau_{B^+}$ is the $B^+$-meson lifetime. Note that terms proportional to the mass of dark antibaryon in (\ref{eq:widthb}) originate from the terms proportional to $\slashed q$ in the decomposition (\ref{eq:ff}). The corresponding expression for model $(b)$ is obtained by the replacement $(d)\to(b)$  in (\ref{eq:widthb}).

\begin{figure}[t]
    \centering
    \includegraphics[width=0.48\textwidth]{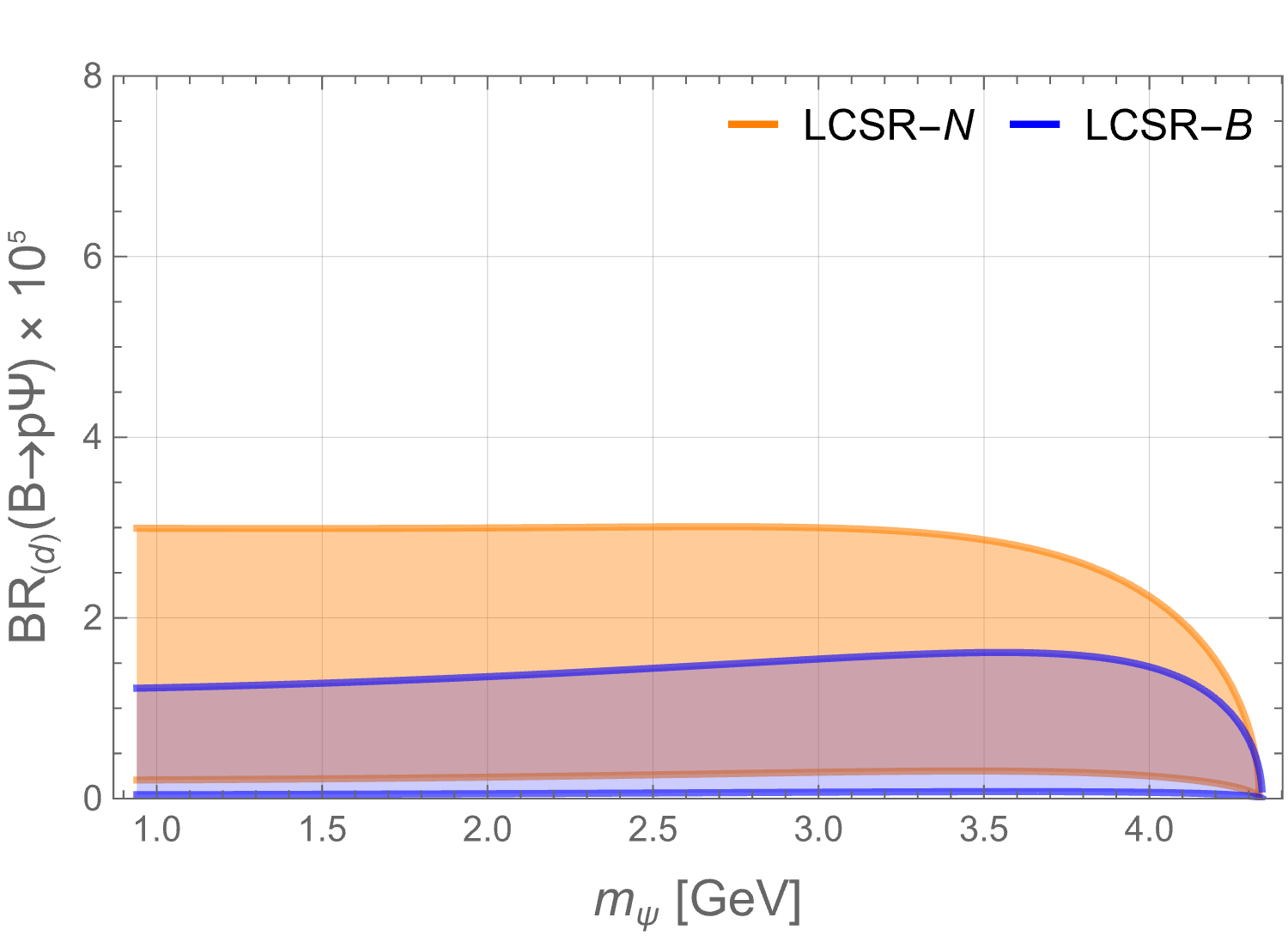
}
    \hfill
    \includegraphics[width=0.48\textwidth]{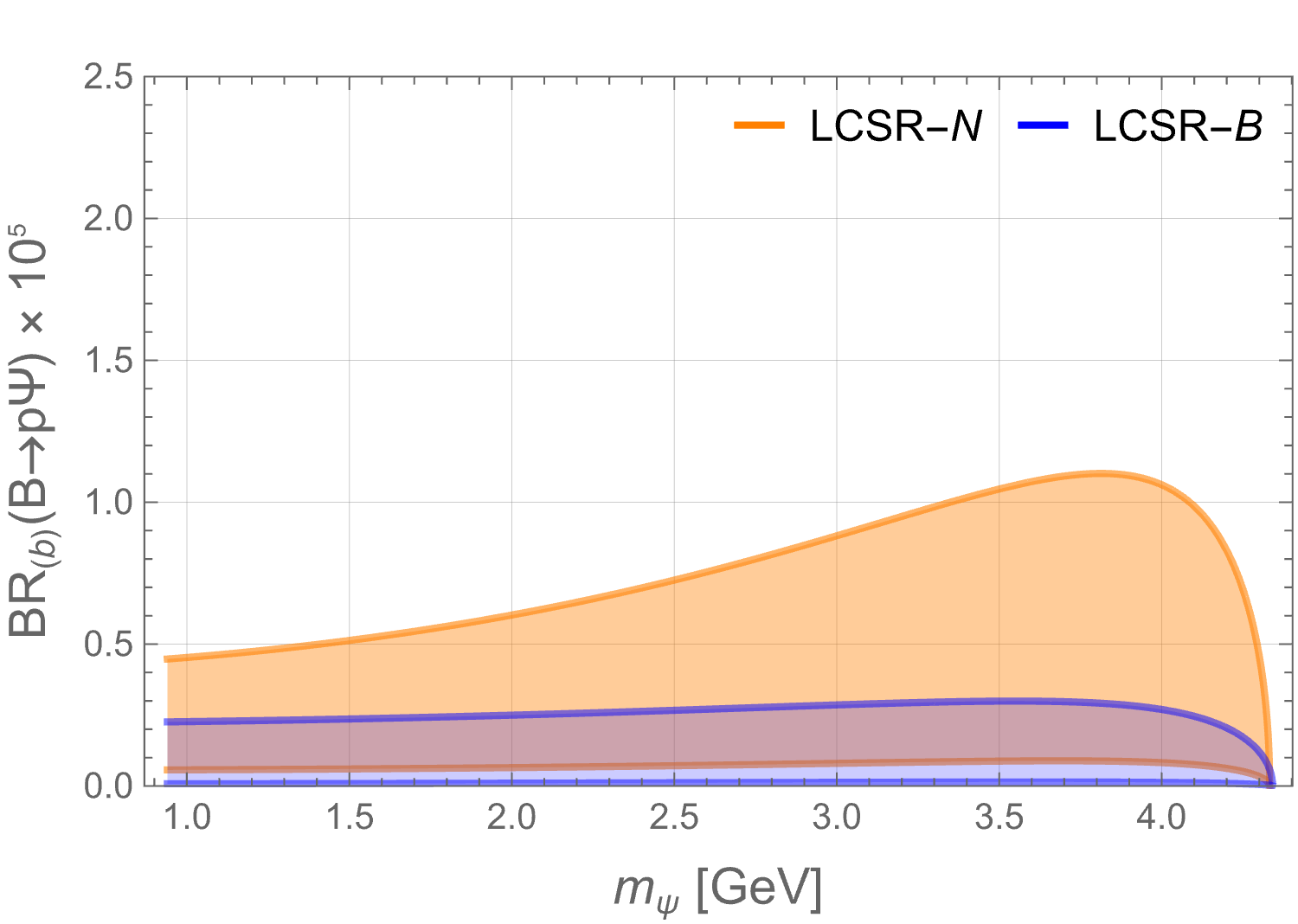}
    \caption{Branching fraction predicted 
    for the $B^+ \to p \Psi$ decay in the  model $(d)$ (left panel) and model $(b)$ (right panel). The blue bands represent the
     results of this work obtained with the LCSR-$B$ method.
     For comparison, the orange bands show the results obtained using the method with nucleon DAs (LCSR-$N$).}
    \label{fig:BR_comparisons}
\end{figure}

The uncertainty of the branching fraction is estimated by independently varying the $z$-expansion parameters (the normalization and slope) of each form factor entering~(\ref{eq:widthb}) over the ranges listed in Table~\ref{tab:FitParameters}, using three representative values for each parameter: central, maximal, and minimal. For every value of $m_\Psi$, the branching fraction is evaluated over the full parameter space, and the uncertainty is defined by the global minimum and maximum obtained in this sampling procedure.
The resulting branching fractions for models $(d)$ and $(b)$ are shown in Fig.~\ref{fig:BR_comparisons}, where the uncertainty band is formed as explained above.

We compare our predictions with the ones in the 
LCSR-$N$ approach \cite{Boushmelev:2023huu}. For both results 
displayed in this figure, the effective $\Psi$-triquark couplings in the 
mesogenesis models  $(d)$ and $(b)$ are set,
for definiteness, to their maximum allowed values 
$|G_{(d)}|^2 = 1.0\times 10^{-13} \,\text{GeV}^{-4}, \, |G_{(b)}|^2 = 4.7 \times10^{-15} \,\text{GeV}^{-4}$. These values 
\footnote{Note that in~\cite{Khodjamirian:2022vta,Boushmelev:2023huu}, equal couplings $|G_{(d)}|^2 = |G_{(b)}|^2 = 10^{-13}$ GeV$^{-4}$ were used.}  are 
equal to the upper limits, inferred, as explained  in~\cite{Alonso-Alvarez:2021qfd}, from the 
ATLAS and CMS searches   for heavy colored particles. 
The results of both LCSR approaches agree within 
large uncertainties. At the same time our
predictions are systematically smaller and also have smaller uncertainties. 

For numerical illustration, we quote the branching fractions obtained with our LCSR-$B$ method at the benchmark value
$m_\Psi = 2~\mathrm{GeV}$ and normalised to the corresponding coupling constants:
\begin{eqnarray}
&&\mathrm{BR}_{(d)}(B^+\to p \Psi)=
\left(2.9^{+10.0}_{-2.4}\right)
\left(\frac{|G_{(d)}|^2}{1.0\times 10^{-13}}\right)
\times 10^{-6}, 
\nonumber\\
&&\mathrm{BR}_{(b)}(B^+\to p\Psi)
=\left(0.54^{+2.0}_{-0.44}\right)
\left(\frac{|G_{(b)}|^2}{4.7\times10^{-15}}\right)
\times 10^{-6}.
\label{eq:exclbenchm}
\end{eqnarray}

\section{Ratio of the $B^+\to p \Psi$ width to inclusive width }
\label{sect:excl-incl}

Our predictions for  the 
partial width of $B^+\to p\Psi $ are not yet sufficient 
for a direct comparison with upper bounds measured for the 
$B^+\to p +invisible$ decay. Indeed, the width is proportional to the square of the effective coupling 
$G_{(d)}$ or $G_{(b)}$, inherent to the model of mesogenesis (see~(\ref{eq:widthb})). For these couplings  only 
the upper limits mentioned above are known. To enable a comparison 
with experiment independently of the coupling, it was
suggested in~\cite{Lenz:2024rwi} to consider the quantity 
\begin{equation}
R_{[excl/incl]}(B^+ \to p\Psi) \equiv
\frac{\mathrm{BR}(B^+ \to p\Psi)}{\mathrm{BR}(B^+\to X_N \Psi)} \times10^{-4} \, ,
\label{eq:Rdef}
\end{equation}
where the denominator contains the branching fraction of the inclusive decay 
$B\to X_N \Psi$. Here  $X_N$ denotes  all  possible and kinematically accessible 
hadronic states  formed after the quark-level decay $\bar{b}\to ud \Psi$ by the diquark $ud$ 
and by the spectator $u$-quark of $B^+$-meson. The ratio of exclusive to 
inclusive widths in (\ref{eq:Rdef}) is multiplied by 
a factor $10^{-4}$, which is the minimal value 
of the inclusive branching fraction 
required for the realization of $B$-mesogenesis~\cite{Alonso-Alvarez:2021qfd}.  
Evidently, in this ratio the effective couplings cancel each other.
The idea put forward in \cite{Lenz:2024rwi} is to evaluate -- at a given $\Psi$ mass
and within a certain model of mesogenesis --
the ratio of exclusive to inclusive widths, using available QCD-based methods for both widths. 
Then the quantity  (\ref{eq:Rdef}) directly yields a lower limit for the exclusive width:
\be
\mathrm{BR}(B^+ \to p\Psi) >R_{[excl/incl]}(B^+ \to p\Psi). 
\label{eq:lowlim}
\ee
Comparing  this limit with  the measured upper bound,  it is possible to determine
the still allowed and/or already excluded values  of the $\Psi$ mass.

In \cite{Lenz:2024rwi}, the inclusive rate 
$\mathrm{BR}(B^+\to X_N \Psi)$ was computed within the framework of the 
heavy quark expansion (HQE), where only the leading dimension-three term was included.  
The analysis was later improved in~\cite{Mohamed:2025zgx}, 
by incorporating the power-suppressed contributions to the inclusive width up to dimension-six level.  
In both studies, the exclusive width entering the ratio was taken from~\cite{Boushmelev:2023huu},  where LCSRs based on nucleon DAs were used.  

Here, we revisit the estimate of $R_{[excl/incl]}$, using our updated 
exclusive widths obtained from LCSRs with $B$-meson DAs. Following~\cite{Mohamed:2025zgx}, we include in the computation of the  inclusive width 
the leading dimension-three term of HQE, as well as  power-suppressed terms 
stemming from the
dimension-five, dimension-six two-quark operators (the Darwin term), and also from the dimension-six 
four-quark 
operators. The detailed analysis of   higher-dimensional terms  in~\cite{Mohamed:2025zgx}  reveals that
the HQE remains reliable up to $m_\Psi \approx 2.0$ GeV for the model~$(d)$ 
and up to $m_\Psi \approx 3.0$ GeV for the model~$(b)$. Above these masses,  subleading
contributions to HQE become dominant and the whole expansion loses convergence. 

\begin{figure}[t]
    \centering
    \includegraphics[width=0.49\textwidth]{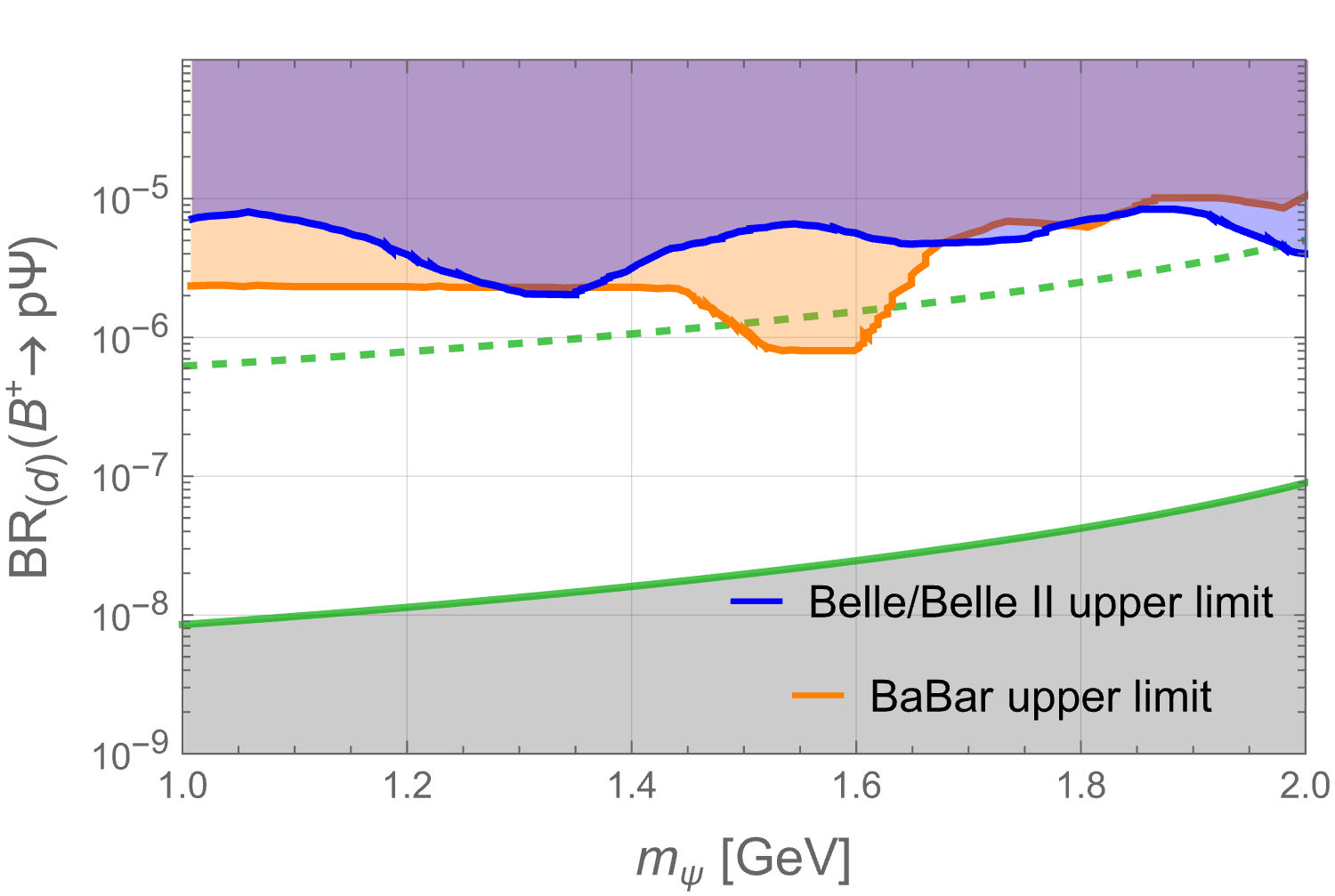}
    \hfill
    \includegraphics[width=0.49\textwidth]{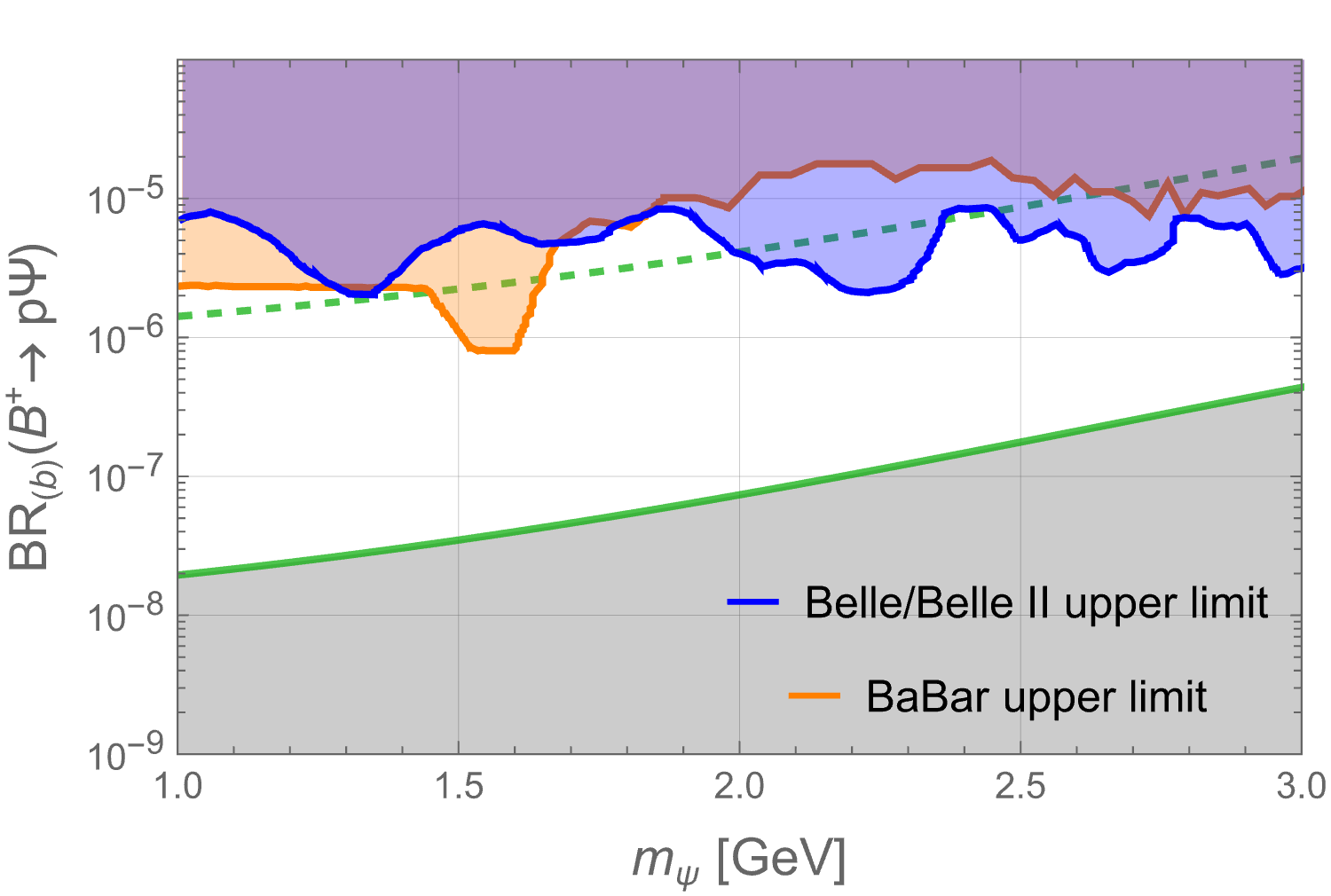}
    
\caption{Branching fraction $\mathrm{BR}(B^+ \to p \Psi)$ as a function of $m_\Psi$ for model $(d)$ (left panel) and model $(b)$ (right panel). The solid green lines present the predicted lower limit  (\ref{eq:lowlim}), and the dashed green lines 
indicate maximal value of this limit within estimated uncertainties; orange and blue shaded regions display  experimental upper limits set by, respectively,
BaBar~\cite{BaBar:2023dtq} and Belle/Belle II~\cite{Belle-II:2026tyb} measurements of 
the $B\to p+invisible$  decays.
The white areas indicate the parameter space allowed by current experimental data. 
}
\label{fig:Rkh}
\end{figure}

Our results for models $(d)$ and $(b)$ are presented in Fig.~\ref{fig:Rkh}, which shows the predicted lower limits for $\mathrm{BR}(B^+ \to p \psi)$ as a function of the dark antibaryon mass $m_\Psi$, together with experimental upper bounds from BaBar~\cite{BaBar:2023dtq} and Belle/Belle~II~\cite{Belle-II:2026tyb}. Our new LCSR-$B$ predictions for the  exclusive branching fractions are smaller than the LCSR-$N$ ones used
in \cite{Lenz:2024rwi}, leading to correspondingly smaller lower limits. In model $(d)$, the lower limits can reach the level of $\sim 10^{-8}$, whereas model $(b)$ yields limits of order $10^{-7}$.
 As explained above, the 
intervals of $\Psi$ mass above 2 GeV (3 GeV) for model $(d)$ (model $(b)$) are not accessible to our 
estimate, due to inapplicability of the HQE method for the inclusive width.

The uncertainty of  the predicted lower limit—represented by the band between the dashed and solid green lines in Fig.~\ref{fig:Rkh}—is determined through the same procedure as employed for the exclusive branching fractions and described in Sec.~\ref{sect:BR}. 
An additional $10\%$ is assigned to account for residual theoretical uncertainty in the inclusive width calculation, including renormalization-scale dependence, scheme choices, and neglected higher-order perturbative and power corrections. Note that the overall uncertainty of the branching fraction is dominated by the form factors $F^{(d)}_{B\to p_R}(0)$ and $F^{(b)}_{B\to p_R}(0)$ computed in Sec.~\ref{sect:num}. Conservatively, we interpret the minimal value of the uncertainty band as our prediction
for the lower limit of $\mathrm{BR}(B^+\to p \Psi)$.

\section{Conclusion}
\label{sect:concl}

In this paper, we applied the method of QCD light-cone sum rules (LCSRs) with $B$-meson distribution amplitudes (DAs) 
to the  $B\to p $ transition form factors.
These hadronic quantities 
are needed to evaluate the width of the exclusive $B^+\to p\Psi$ decay 
predicted in the $B$-mesogenesis models. We considered two such models, 
$(d)$ and $(b)$, with different $\Psi$-triquark effective interactions.

LCSRs for these unusual baryon-to-meson form factors 
were obtained, expanding the underlying $B\to$ vacuum correlation function in terms of $B$-meson DAs, up  to the twist-5 accuracy. 
 We found that this correlation function has  quite a peculiar operator-product expansion (OPE), as compared to traditional applications of this method to the $B\to meson$ form factors.
First of all, the leading-order diagram  contains a diquark loop, leading to logarithmic terms in the OPE. 
Moreover, the 
 contributions of all three-particle (quark-antiquark-gluon) DAs up to twist-4  vanish. That is caused by  several reasons: 
the form of the proton interpolating current, the scalar structure of the underlying effective operator, and the adopted limit of massless $u,d$ quarks.
We  also found that the  form factors 
in the two mesogenesis models are simply related by a factor of two.  
Altogether, the LCSRs  for $B\to p$ form factors obey the hierarchy
of twists, so that power corrections stemming from the subleading terms of light-cone expansion 
and expressed via twist-4,5 DAs 
are numerically suppressed, albeit moderately, with respect to the leading contributions from the twist-2,3 DAs. 

For the hadronic part of our sum rules, describing 
the  channel of the proton interpolating current,
we adopted the same  ansatz as in the LCSRs for the nucleon electromagnetic  form factors, in which all  excited nucleon states are included in the continuum and approximated by quark-hadron duality.

For the two $B\to p$ form factors relevant for the $B\to p \Psi$ decay width, we produced numerical 
results in the region of the LCSR validity at negative momentum transfer squared. The form factors in the physical region of that  decay were then evaluated by fitting the LCSR results to a standard $z$-expansion and by the subsequent extrapolation of the latter. 

Our results for $B\to p$ form factors are compared  with
the ones from the alternative method of LCSRs with nucleon DAs
(LCSR-$N$). The latter sum rules, especially the ones for the model
$(b)$, reveal an anomalously large higher-twist contributions 
with still uncertain  parameters, 
making the LCSR-$N$ method somewhat less reliable than the LCSR-$B$ method applied in this paper. Nevertheless, the widths predicted from the two approaches and for both models $(d)$ and $(b)$ agree within large uncertainties. Numerically, the range of $B^+\to p\Psi$ branching fractions obtained  in this work is smaller and narrower.

Apart from 
predictions for the exclusive $B^+\to p\Psi$ decay width as a function of the $\Psi$ mass and effective coupling, we 
evaluated  the important ratio of  exclusive to inclusive 
branching fractions, employing the available heavy quark expansion (HQE) result for inclusive widths. This ratio
yields lower bounds for the  $B^+\to p\Psi$ branching fraction,
independent of the 
value of the effective $\Psi$-triquark coupling.
Comparison with existing experimental upper bounds reveals that there is still a large gap between upper and lower bounds in the region of small $\Psi$ masses.

Further improvements and extensions of LCSRs obtained in this paper are possible. In particular, more accurate estimates of the two key parameters: the inverse moment of the $B$-meson twist-2 DA and the nucleon decay constant, will decrease the parametric uncertainty of LCSRs. Furthermore, a more detailed resonance pattern of the hadronic spectral density  in the sum rule will open up a possibility to estimate also the form factors of the $B\to N^*(1535)$ and $B\to N(1440)$ transitions. 

Most importantly, the LCSRs with $B$-meson DAs can be straightforwardly extended to evaluate  the form factors of $B\to \Lambda$ or $B\to \Lambda_c$ transitions, using the interpolating three-quark current with a suitable quark-flavour combination
 \footnote{
After this work was completed, 
we became aware of the very recent study
\cite{Hiller:2026osz}, where searches for jets with missing energy at the
LHC are used to constrain baryon-number violating operators relevant for
$B$-mesogenesis. In particular, the quark-flavour combination $bud$ appears to be disfavored, while scenarios involving
heavier flavour structures such as $bcs$ remain viable, leading to $B^{0,+}\to \Xi_c^{0,+}\Psi$
and $B_s\to \Omega_c^0\Psi$ decay modes. This observation further motivates the use of flavour-flexible LCSR framework
suggested in this work.
}.
This framework can therefore be applied to the $B\to \Lambda \Psi$ and $B^+\to \Lambda^+_c \Psi$ decay modes predicted in various flavour realizations of $B$-mesogenesis models.
In this context it is worth mentioning also the
recent application \cite{Shi:2024uqs} of LCSRs with the $\Lambda_b$  DAs,  introduced in \cite{Ball:2008fw}, to the form factors of $\Lambda_b\to \pi,K $  transitions, relevant for the $\Lambda_b\to \pi\Psi$ and $\Lambda\to K \Psi$ decays.

In the absence of lattice QCD calculations of $B\to p$ form factors,  LCSRs remain the only QCD-based method to provide hadronic input
to the studies of $B^+\to p\Psi$ and similar invisible $B$-meson decays. Our results obtained from these sum rules indicate that future progress in two directions -- reducing the uncertainty of the theoretical lower limits and decreasing the experimental upper limits by at least an order of magnitude -- could provide a decisive test of the $B$-mesogenesis scenario.

\subsection*{Acknowledgments}
We thank Alexander Lenz for useful and stimulating discussions.
This work  was supported by the Deutsche Forschungsgemeinschaft 
(DFG, German Research Foundation) under the grant 396021762 - TRR 257.

\appendix 
\numberwithin{equation}{section}

\section{$B$-meson distribution amplitudes}
\label{sect:appBDA}
\subsection{Two-particle DAs}
We use the following decomposition 
\cite{Grozin:1996pq}
of the $B^+$-to-vacuum matrix element in HQET: 
\begin{align}
\langle 0|  \bar{h}_{v\alpha}(0)[0,x] \, u_\beta(x) | B^+(v)\rangle 
=\frac{i f_B m_B}4 
\int\limits_0^\infty\!\! d\omega \, e^{-i\omega v\cdot x}
\Bigg[
\Big\{\phi^B_+(\omega) + x^2 g_+^B(\omega)\Big\}
\Big[(1+\slashed v)\gamma_5  \Big]_{\beta\alpha}
\nonumber\\
- \frac{1}{2\, v\cdot x}
\Big\{\phi^B_+(\omega)-\phi^B_-(\omega) + 
x^2 \big(g_+^B(\omega)-g_-^B(\omega)\big)\Big\} 
\Big[\slashed{x} (1+\slashed v) \gamma_5 \Big]_{\beta\alpha} \Bigg]\,.
\label{eq:tw4}
\end{align}
In the above, $[0,x]$  is the gauge link,
and the variable $\omega$ is the light-cone projection of the light-quark momentum in the heavy-meson rest frame.

The functions $\phi^B_+(\omega)$, $\phi^B_-(\omega)$, $g^B_+(\omega)$ and $g^B_-(\omega)$ are, respectively, the DAs of twist two, three, four and five, according to the nomenclature of \cite{Braun:2017liq}. The normalization conditions for the first two DAs are:
\begin{equation}
    \int\limits_{0}^{\infty}\!\! d\omega \,\phi_{\pm}^{B}(\omega) = 1\,. 
\label{eq:DANorm}
\end{equation}
The decomposition~(\ref{eq:tw4}) is rearranged into an equivalent form without the factors $1/(v\cdot x)$, by introducing the integrated linear combination of the twist-2 and twist-3 DAs~\cite{Khodjamirian:2006st}: 
\begin{align}
    \Phi^B_{\pm}(w)  \equiv \int\limits_0^w \!d\tau \left(\phi^B_+(\tau) - \phi^B_-(\tau)\right), 
\label{eq:Phi}
\end{align}
so that 
\begin{align}
\frac{d\Phi^B_{\pm}(\omega) }{d \omega} = \phi^B_{+}(\omega) - \phi^B_{-}(\omega)\,,~~
\Phi^B_{\pm}(0) =  \Phi^B_{\pm}(\infty) = 0. ~~
\end{align}
We also use analogous definition for the 
combination of twist-four and twist-five DAs:
\begin{align}
G^B_{\pm}(\omega) \equiv \int\limits_0^ \omega \!d\tau \,\left(g^B_+(\tau)-g^B_-(\tau)\right). 
\label{eq:Gplus}
\end{align}
Note that
performing charge conjugation of the matrix element in (\ref{eq:tw4}), we restore for the twist-two and twist-three parts the definition of $B^-$-meson DAs
used e.g, in \cite{Beneke:2000wa,Khodjamirian:2006st}. 

In this paper we use the standard 
exponential model \cite{Grozin:1996pq} of the twist-two DA:
\begin{equation}
\phi^B_+(\omega) = 
\frac{\omega}{\lambda_B^2} e^{-\omega/\lambda_B}\,.
\label{eq:phiBplus}
\end{equation}
For the twist-three and twist-four DAs we emlpoy the corresponding exponential form
suggested in \cite{Braun:2017liq}:
\begin{eqnarray}
&&\phi^B_-(\omega) = \frac1{\lambda_B} e^{-\omega/\lambda_B}
+ \frac1{2\lambda_B} \bigg[ 1 - \frac{2\omega}{\lambda_B} + \frac{\omega^2}{2\lambda_B^2} \bigg] \frac{1-R}{1+2R}\, e^{-\omega/\lambda_B} \,,
\label{eq:phiBminus}\\
&&g^B_+(\omega) = \frac{3 \omega^2}{16\lambda_B} \frac{3+4R}{1+2R} \,e^{-\omega/\lambda_B}\,.
\label{eq:gBplus}
\end{eqnarray}
Finally, for the twist-five DA
we adopt the same exponential form  based on Wandzura-Wilczek approximation, as in \cite{Gubernari:2018wyi}:
\be
g^B_{-}(\omega)= \frac{3\omega}{4}e^{-\omega/\lambda_B}.
\label{eq:gmin}
\ee

To assess the role of twist-five contribution
in the numerical analysis, we also omit the term
with $g^B_{-}$ in the correlation function. This step is reduced to a simple replacement: 
\be
G^B_{\pm}(\omega)\to G^B_{+}(\omega)=\int\limits_0^ \omega \!d\tau \,g^B_+(\tau) 
\label{eq:Gplmin}
\ee
in (\ref{eq:imF1}) and (\ref{eq:imF4}).

The set of $B$-meson DAs used here is determined by three  parameters. The first one is  the $B$-meson decay constant, very accurately computed in lattice QCD; the second one is the inverse moment of the twist-2 DA:
\begin{equation}
\frac{1}{\lambda_B} = \int\limits_0^\infty d\omega 
\frac{\phi^B_+(\omega)}{\omega}\,.
\label{eq:lamB}
\end{equation}
The adopted value quoted in 
Table~\ref{tab:input}  was obtained 
\cite{Braun:2003wx}
at the scale $\sim 1$ GeV,
and here we neglect the scale dependence for simplicity, having in mind the currently large uncertainty of the available determination. The third parameter is the ratio
\be
R\equiv\lambda_E^2/\lambda_H^2\,,
\label{eq:R}
\ee
where $\lambda_{E,H}$ 
parameterize the vacuum-to-$B$ matrix elements of the 
heavy-light quark-antiquark-gluon operators.
Their definitions and the first estimates can be 
found in \cite{Grozin:1996pq}. These estimates 
were obtained, employing QCD sum rules in HQET for a nondiagonal
correlation function, in which the product of a quark-antiquark-gluon operator is 
correlated with the $B$-meson interpolating quark-antiquark current. 
These determinations were improved
in two different directions: in  \cite{Nishikawa:2011qk}, adding NLO corrections to the nondiagonal sum rules and in  \cite{Rahimi:2020zzo}, employing 
diagonal correlation functions. The results only marginally agree, therefore
we take an average of both.
As inputs for averaging we take  the three intervals given by 
$\lambda^2_E(1~\mbox{GeV})= 0.03\pm 0.02$ GeV$^2$,   $\lambda^2_H (1~\mbox{GeV})= 0.06\pm 0.03$ GeV$^2$ from  \cite{Nishikawa:2011qk}
and  $R=0.1\pm 0.1$ from \cite{Rahimi:2020zzo}.
Assuming all these intervals are uniformly 
distributed and uncorrelated, we used a large $(10^5)$ sample of points from 
the first two of them, and obtain a distribution for $R$
calculated from (\ref{eq:R}). A similar sample for this parameter was also drawn directly from the third interval.
The resulting range   for $R$ quoted in Table~\ref{tab:input}  was inferred from the distribution of the arithmetic mean of these samples, with the central value corresponding to the median and the uncertainties corresponding to the $68.27\%$ $(1\sigma)$ deviation.
Our estimate is consistent   with previous one  $R=0.4^{+0.5}_{-0.3}$ from \cite{Descotes-Genon:2019bud}   determined differently, using as inputs  the sum rule results from \cite{Grozin:1996pq} and \cite{Nishikawa:2011qk}.

\subsection{Three-particle DAs }

We use the $B\to vacuum$ matrix element expanded in terms of 
three-particle (quark-antiquark-gluon)  DAs with twist-4 accuracy 
that was originally obtained in \cite{Braun:2017liq}.
More specifically, we use the expression  of this expansion
for $\bar{B}$ meson from  \cite{Descotes-Genon:2019bud},
rewriting it with free Dirac indices and charge-conjugating to the
matrix element with $B^+$ meson:

\begin{eqnarray}
&&
\lefteqn{\langle 0| \bar h_{v\alpha}(0) 
 g_s \frac{\lambda^a}2 G^a_{\mu\nu}(ux) q_\beta(x)| B^+(v)\rangle =
-\frac{f_B m_B}4 \int_0^\infty d\omega_1 \int_0^\infty d\omega_2\ e^{-i (\omega_1 + u \omega_2) v\cdot x}}
\nonumber\\
&&\times \biggl\{  \gamma_5 \biggl[
-(v_\mu\gamma_\nu-v_\nu\gamma_\mu)  \phi_3
-\frac{i}2\sigma_{\mu\nu} \big[ \phi_3 - \phi_4 \big]
+ \frac{x_\mu v_\nu-x_\nu v_\mu}{2v\cdot x} \big[ \phi_3 + \phi_4 - 2\psi_4 \big]
\nonumber\\
&&+\frac{x_\mu \gamma_\nu-x_\nu \gamma_\mu}{2v\cdot x}\big[\phi_3 + \widetilde\psi_4 \big]
+ \frac{i\epsilon_{\mu\nu\alpha\beta} x^\alpha v^\beta}{2v\cdot x}  \gamma_5\,
\big[ \phi_3 - \phi_4 +2 \widetilde\psi_4 \big]
\nonumber\\
&&- \frac{i\epsilon_{\mu\nu\alpha\beta} x^\alpha}{2v\cdot x}  \gamma^\beta\gamma_5\, \big[ \phi_3 - \phi_4 + \widetilde\psi_4 \big]
+\frac{(x_\mu v_\nu-x_\nu v_\mu)\slashed{x}}{2(v\cdot x)^2} \, \big[ \phi_4 - \psi_4 - \widetilde\psi_4 \big]
\nonumber\\
&&- \frac{ \slashed{x}  (x_\mu \gamma_\nu-x_\nu \gamma_\mu)}{4(v\cdot x)^2} \, \big[ \phi_3 - \phi_4 + 2 \widetilde\psi_4 \big]
\biggr](1-\slashed{v})\biggr\}_{\beta\alpha}
\,.
\label{eq:def3DAsBpl}
\end{eqnarray}
In the above, $\phi_3$ is the DA of twist-3,   
and the three DAs of twist-4 are 
$\phi_4,\psi_4$ and $\widetilde\psi_4 $.
For brevity, their dependence  on $\omega_1,\omega_2$ is 
not shown, as well as the  Wilson lines on left-hand side.

In analogy to (\ref{eq:Phi}) and 
(\ref{eq:Gplus}),  we define the integrated DAs,
\be
\overline \Phi (\omega_1,\omega_2) = \int_0^{\omega_1} d\tau\,\Phi (\tau,\omega_2)\ , \qquad
\label{eq:barPhi}
\ee
where $\Phi = \{ \phi_3,\phi_4,\psi_4, \widetilde \psi_4 \}$.
This time we repeat this procedure twice to 
get rid of the second power of $(1/v\cdot x)$
in (\ref{eq:def3DAsBpl}), introducing DAs
with double bars: 
\be
\overline{\overline \Phi} (\omega_1,\omega_2) = \int_0^{\omega_1} d\tau\, \overline\Phi (\tau,\omega_2).
\label{eq:doublebarPhi}
\ee
Applying (\ref{eq:barPhi}) and (\ref{eq:doublebarPhi}), it is possible,
--  as it was done in~\cite{Descotes-Genon:2019bud} -- 
to transform  the equation (\ref{eq:def3DAsBpl}) 
to a more convenient form, free of $(v.x)$ in denominators:
\ba
&&\langle 0| \bar h_{v\alpha}(0) 
 g_s \frac{\lambda^a}2 G^a_{\mu\nu}(ux) q_\beta(x)| B^+(v)\rangle
= -\frac{f_B m_B}4 \int_0^\infty d\omega_1\, 
d\omega_2\ e^{-i \sigma m_B v\cdot x} \, 
\nonumber\\
&&\times\biggl\{ \gamma_5\biggl[
-(v_\mu\gamma_\nu-v_\nu\gamma_\mu)  \phi_3
-\frac{i}2\sigma_{\mu\nu} \big[ \phi_3 - \phi_4 \big]
+ \frac{i}2 (x_\mu v_\nu-x_\nu v_\mu) \big[ \overline \phi_3 + \overline \phi_4 - 2 \overline \psi_4 \big]
\nonumber\\
&&+\frac{i}2 (x_\mu \gamma_\nu-x_\nu \gamma_\mu) \big[ \overline \phi_3 + \overline{\widetilde\psi}_4 \big]
- \frac12 \epsilon_{\mu\nu\alpha\beta} x^\alpha v^\beta  \gamma_5\,
\big[ \overline \phi_3 - \overline \phi_4 +2 \overline {\widetilde\psi}_4 \big]
\nonumber\\
&&+ \frac12 \epsilon_{\mu\nu\alpha\beta} x^\alpha  \gamma^\beta\gamma_5\, \big[ \overline\phi_3 - \overline\phi_4 + \overline{\widetilde\psi}_4 \big]
- \frac12 (x_\mu v_\nu-x_\nu v_\mu)\slashed{x} \, \big[ \overline{\overline\phi}_4 -  \overline{\overline\psi}_4 - \overline{\overline{\widetilde\psi}}_4 \big]
\nonumber\\
&&+ \frac14  \slashed{x} (x_\mu \gamma_\nu-x_\nu \gamma_\mu) \, \big[ \overline{\overline\phi}_3 - \overline{\overline\phi}_4 + 2 \overline{\overline{\widetilde\psi}}_4 \big]
\biggr](1-\slashed{v})
\biggr\}_{\beta\alpha}
\,,
\label{eq:3DA_compact}
\ea
where we introduce the new variable
\be
\sigma(u,\omega_1,\omega_2) \equiv (\omega_1 + u \,\omega_2)/m_B\,.
\label{eq:sigma}
\ee

The exponential model (\ref{eq:phiBplus}) was extended to
the three-particle $B$-meson DAs in~\cite{Braun:2017liq}, and we present that model here for completeness, although we will not use them
due to vanishing of three-particle DAs contributions
to our correlation function (see Sec.~\ref{sect:ope_3DAs}).
 We have:
\ba
\phi_3(\omega_1,\omega_2) &=&  - \frac{3\,\omega_1 \omega_2^2}{4 \lambda_B^3} \frac{1-R}{1+2R}  \,e^{- (\omega_1+\omega_2)/\lambda_B} 
\nonumber \\
\phi_4(\omega_1,\omega_2) &=&  \frac{3\,\omega_2^2}{4 \lambda_B^2} \frac{1+R}{1+2R} 
\,e^{- (\omega_1+\omega_2)/\lambda_B} 
\nonumber\\
\psi_4(\omega_1,\omega_2) &=& R \ \widetilde \psi_4(\omega_1,\omega_2)
= \frac{3\,\omega_1 \omega_2}{2 \lambda_B^2} \frac{R}{1+2R} \,e^{-(\omega_1+\omega_2)/\lambda_B} \,.
\label{eq:3DAsexpmod}
\ea
Note that all three-particle DAs in this model are characterized by
two input parameters, $\lambda_B$ and $R$, the same  as in the two-particle DAs.
 
Finally, we present the coefficients 
in the expression (\ref{eq:3partdiag}) for  one of the diagrams 
with three-particle DAs:
\begin{align}
 {\cal C}(\mathcal{P},v) &= \big[8(\mathcal{P} \cdot v)\slashed{\mathcal{P}}\slashed{v} - 16(\mathcal{P} \cdot v)^2\big] \phi_3 + 8(\mathcal{P} \cdot v - \slashed{\mathcal{P}}\slashed{v})(\overline{\phi}_3 - \overline{\phi}_4) + 8(4\mathcal{P} \cdot v - \slashed{\mathcal{P}}\slashed{v}) \overline{\tilde{\psi}}_4 
 \nonumber\\
&\quad - \frac{4}{\mathcal{P}^2}[\mathcal{P}^2 + 2(\mathcal{P} \cdot v)\slashed{\mathcal{P}}\slashed{v}](\overline{\overline{\phi}}_4 - \overline{\overline{\psi}}_4 - \overline{\overline{\tilde{\psi}}}_4) - 12(\overline{\overline{\phi}}_3 - \overline{\overline{\phi}}_4 + 2\overline{\overline{\tilde{\psi}}}_4), 
\nonumber\\
\overline{\cal{C}}(\mathcal{P},v) &= 12 \, \mathcal{P}^2 \,\text{Log}(-\mathcal{P}^2) \phi_4 - 4(\mathcal{P} \cdot v - \slashed{\mathcal{P}}\slashed{v})(\overline{\phi}_3 - \overline{\phi}_4) - 4(4\mathcal{P} \cdot v - \slashed{\mathcal{P}}\slashed{v}) \overline{\tilde{\psi}}_4 
\nonumber \\
&\quad + \frac{4}{\mathcal{P}^2}[\mathcal{P}^2 + 2(\mathcal{P} \cdot v)\slashed{\mathcal{P}}\slashed{v}](\overline{\overline{\phi}}_4 - \overline{\overline{\psi}}_4 - \overline{\overline{\tilde{\psi}}}_4) + 12(\overline{\overline{\phi}}_3 - \overline{\overline{\phi}}_4 + 2\overline{\overline{\tilde{\psi}}}_4),
\label{eq:3partdiag1}
\end{align}
where ${\cal P} \equiv P-
\sigma m_B v$. In these expressions, terms without an imaginary part in $P^2$ are omitted since they do not contribute to the LCSRs.
Note however, that the relations (\ref{eq:3partrel1}) 
and (\ref{eq:3partrel2}) for three-particle contributions are valid including these terms.

\section{Transformation to a form of dispersion integral}
\label{app:disp}
To transform the amplitudes 
(\ref{eq:corrF1}) and (\ref{eq:corrF4}) into a dispersion form, 
we subdivide their expressions into similar terms
with respect to analytic properties in the variable 
$P^2$.

For the terms that contain $s(w)-P^2$ in denominator
the transformation is most easily done. 
For each term of this type  multiplied by 
a certain function $g(\omega)$ we use the transformation 
$\omega\to s(\omega)$ and obtain:
\be
\int\limits_{0}^{\infty}\! d\omega 
\,\frac{g(\omega)}{s(\omega)-P^2}
= \int\limits_{0}^{\infty}\! 
\frac{ds}{s-P^2}\,
\frac{d\omega(s)}{ds}g(\omega(s))
\,.
\label{eq:intdenom}
\ee

To obtain a dispersion form for the terms 
in (\ref{eq:corrF1}) and (\ref{eq:corrF4})
with a logarithmic dependence on $P^2$, it is sufficient to consider the integral:
\be
I_{(f)}(P^2)\equiv \int\limits_{0}^{\infty}\! d\omega 
\ln\!\big[ s(\omega)-P^2\big]
f(\omega)= \int\limits_{0}^{\infty}\! ds
\ln\!\big[ s-P^2\big] \frac{d\omega(s)}{ds}
f(\omega(s))\,,
\label{eq:intLog}
\ee
where $f(\omega)$ is a function 
containing one of the DAs. 
Employing $$\frac{1}{\pi}\mbox{Im}_{P^2}\big(\ln\!\big[ s-P^2\big]\big)= -\theta(P^2-s)\,,$$
and returning  in (\ref{eq:intLog})
to the initial integration variable $\omega$, we obtain:
\be
\frac{1}{\pi}\mbox{Im} I_{(f)}(s)=-\!\!\int\limits_{0}^{\omega(s)}\!\! d\omega' \,f(\omega')\,,
\label{eq:intLogdisp}
\ee
yielding the dispersion form of that integral:
\be
I_{(f)}(P^2) = -\int_0^\infty 
\frac{ds}{s-P^2}\int\limits_{0}^{\omega(s)}\!\! d\omega' \,f(\omega')\,, 
\label{eq:intLog1}
\ee
Note that the integral over $s$ is divergent
on the upper limit,
but that is not relevant for our purposes, because
this divergence will be eliminated by the Borel 
transform $P^2\to M^2$. In other words, the dispersion integral needs subtractions, and since subtraction terms 
are dependent on polynoms in $P^2$, they
vanish after that transform.
For the same reason, the terms in (\ref{eq:corrF1}), containing an extra factor $P^2$ 
multiplying the above integral 
are easily transformed to the dispersion form 
by replacing  $P^2\to s $, and
\be
P^2\,I_{(f)}(P^2) \to -\int_0^\infty ds
\frac{s}{s-P^2}\int\limits_{0}^{\omega(s)}\!\! d\omega' \,f(\omega')\,.
\label{eq:intLogP2}
\ee

Still it is important 
to convince ourselves that after Borel transform
both expressions,
(\ref{eq:intLog1}) and (\ref{eq:intLog}) are equal.
For this purpose, we derive the Borel transform for the logarithmic function entering (\ref{eq:intLog}):
\ba
{\cal B}_{M^2} \bigg( \ln \big[s(\omega)-P^2\big]\bigg)=
\int\limits_0^{s(\omega)}\! ds'\,e^{-s'/M^2}-M^2\,.
\label{eq:logBorel}
\ea
This formula is obtained if one uses (\ref{eq:intLog1})
and the analogous expression for the imaginary part of $\ln[-P^2]$ and then takes the difference of two
dispersion relations, yielding:
\be
\ln\big[s(\omega)-P^2\big]-\ln\big[-P^2 \big]=
\int\limits_0^{s(\omega)}\! \frac{ds'}{s'-P^2}\,.
\ee
The Borel transform of this equation, together with 
the well familiar formula 
$${\cal B}_{M^2} \ln\big[-P^2\big]=-M^2\,,$$
yields then (\ref{eq:logBorel}). 
Applying the latter formula to the integral 
(\ref{eq:intLog}), we obtain
\be
{\cal B}_{M^2} \big(I_{(f)}(P^2)\big)= -M^2\int\limits_{0}^{\infty}\! d\omega\, 
e^{-s(\omega)/M^2} f(\omega)\,.
\label{eq:intLog2}
\ee
It remains to apply the same Borel transform to 
the dispersion form (\ref{eq:intLog1}):
\be
{\cal B}_{M^2} \big(I_{(f)}(P^2)\big)_{disp} = 
-\int\limits_{0}^{\infty} ds\, e^{-s/M^2}
\int_0^{\omega(s)} \!\! d\omega' \,f(\omega')\,. 
\label{eq:intdisp2}
\ee
As a final step of our proof, this equality is reduced to (\ref{eq:intLog2}) by a simple  permutation of the integration variables $s$ and $\omega'$.

\clearpage
\bibliographystyle{JHEP}
\bibliography{references}

@article{Elor:2018twp,
    author = "Elor, Gilly and Escudero, Miguel and Nelson, Ann",
    title = "{Baryogenesis and Dark Matter from $B$ Mesons}",
    eprint = "1810.00880",
    archivePrefix = "arXiv",
    primaryClass = "hep-ph",
    reportNumber = "KCL-18-53, IFIC-18-35",
    doi = "10.1103/PhysRevD.99.035031",
    journal = "Phys. Rev. D",
    volume = "99",
    number = "3",
    pages = "035031",
    year = "2019"
}

@article{Alonso-Alvarez:2021qfd,
    author = "Alonso-\'Alvarez, Gonzalo and Elor, Gilly and Escudero, Miguel",
    title = "{Collider signals of baryogenesis and dark matter from B mesons: A roadmap to discovery}",
    eprint = "2101.02706",
    archivePrefix = "arXiv",
    primaryClass = "hep-ph",
    reportNumber = "TUM-HEP 1299/20",
    doi = "10.1103/PhysRevD.104.035028",
    journal = "Phys. Rev. D",
    volume = "104",
    number = "3",
    pages = "035028",
    year = "2021"
}

@article{Elahi:2021jia,
    author = "Elahi, Fatemeh and Elor, Gilly and McGehee, Robert",
    title = "{Charged B mesogenesis}",
    eprint = "2109.09751",
    archivePrefix = "arXiv",
    primaryClass = "hep-ph",
    reportNumber = "LCTP-21-24, MITP-21-041",
    doi = "10.1103/PhysRevD.105.055024",
    journal = "Phys. Rev. D",
    volume = "105",
    number = "5",
    pages = "055024",
    year = "2022"
}

@article{Alonso-Alvarez:2021oaj,
    author = "Alonso-\'Alvarez, Gonzalo and Elor, Gilly and Escudero, Miguel and Fornal, Bartosz and Grinstein, Benjam\'\i{}n and Martin Camalich, Jorge",
    title = "{Strange physics of dark baryons}",
    eprint = "2111.12712",
    archivePrefix = "arXiv",
    primaryClass = "hep-ph",
    reportNumber = "TUM-HEP-1373/21, MITP-21-060",
    doi = "10.1103/PhysRevD.105.115005",
    journal = "Phys. Rev. D",
    volume = "105",
    number = "11",
    pages = "115005",
    year = "2022"
}

@article{Nelson:2019fln,
    author = "Nelson, Ann E. and Xiao, Huangyu",
    title = "{Baryogenesis from B Meson Oscillations}",
    eprint = "1901.08141",
    archivePrefix = "arXiv",
    primaryClass = "hep-ph",
    doi = "10.1103/PhysRevD.100.075002",
    journal = "Phys. Rev. D",
    volume = "100",
    number = "7",
    pages = "075002",
    year = "2019"
}

@article{Khodjamirian:2005ea,
    author = "Khodjamirian, Alexander and Mannel, Thomas and Offen, Nils",
    title = "{$B$-meson distribution amplitude from the $B \to \pi$ form-factor}",
    eprint = "hep-ph/0504091",
    archivePrefix = "arXiv",
    reportNumber = "SI-HEP-2005-01",
    doi = "10.1016/j.physletb.2005.06.021",
    journal = "Phys. Lett. B",
    volume = "620",
    pages = "52--60",
    year = "2005"
}

@article{Khodjamirian:2006st,
    author = "Khodjamirian, Alexander and Mannel, Thomas and Offen, Nils",
    title = "{Form-factors from light-cone sum rules with $B$-meson distribution amplitudes}",
    eprint = "hep-ph/0611193",
    archivePrefix = "arXiv",
    reportNumber = "SI-HEP-2006-03",
    doi = "10.1103/PhysRevD.75.054013",
    journal = "Phys. Rev. D",
    volume = "75",
    pages = "054013",
    year = "2007"
}

@article{Braun:2017liq,
    author = "Braun, V. M. and Ji, Yao and Manashov, A. N.",
    title = "{Higher-twist $B$-meson Distribution Amplitudes in HQET}",
    eprint = "1703.02446",
    archivePrefix = "arXiv",
    primaryClass = "hep-ph",
    reportNumber = "DESY-17-037",
    doi = "10.1007/JHEP05(2017)022",
    journal = "JHEP",
    volume = "05",
    pages = "022",
    year = "2017"
}

@article{Ioffe:1981kw,
    author = "Ioffe, B. L.",
    title = "{Calculation of Baryon Masses in Quantum Chromodynamics}",
    reportNumber = "ITEP-10-1981, ITEP-92-1981",
    doi = "10.1016/0550-3213(81)90259-5",
    journal = "Nucl. Phys. B",
    volume = "188",
    pages = "317--341",
    year = "1981",
    note = "[Erratum: Nucl.Phys.B 191, 591--592 (1981)]"
}

@article{Grozin:1996pq,
    author = "Grozin, A. G. and Neubert, M.",
    title = "{Asymptotics of heavy meson form-factors}",
    eprint = "hep-ph/9607366",
    archivePrefix = "arXiv",
    reportNumber = "BUDKER-INP-1996-45, BINP-96-45, CERN-TH-96-144",
    doi = "10.1103/PhysRevD.55.272",
    journal = "Phys. Rev. D",
    volume = "55",
    pages = "272--290",
    year = "1997"
}

@article{Ball:2008fw,
    author = "Ball, Patricia and Braun, Vladimir M. and Gardi, Einan",
    title = "{Distribution Amplitudes of the Lambda(b) Baryon in QCD}",
    eprint = "0804.2424",
    archivePrefix = "arXiv",
    primaryClass = "hep-ph",
    reportNumber = "IPPP-08-18, DCPT-08-36, EDINBURGH-2008-15",
    doi = "10.1016/j.physletb.2008.06.004",
    journal = "Phys. Lett. B",
    volume = "665",
    pages = "197--204",
    year = "2008"
}

@article{Beneke:2000wa,
    author = "Beneke, M. and Feldmann, T.",
    title = "{Symmetry breaking corrections to heavy to light B meson form-factors at large recoil}",
    eprint = "hep-ph/0008255",
    archivePrefix = "arXiv",
    reportNumber = "PITHA-00-20",
    doi = "10.1016/S0550-3213(00)00585-X",
    journal = "Nucl. Phys. B",
    volume = "592",
    pages = "3--34",
    year = "2001"
}

@article{Gubernari:2018wyi,
    author = "Gubernari, Nico and Kokulu, Ahmet and van Dyk, Danny",
    title = "{$B\to P$ and $B\to V$ Form Factors from $B$-Meson Light-Cone Sum Rules beyond Leading Twist}",
    eprint = "1811.00983",
    archivePrefix = "arXiv",
    primaryClass = "hep-ph",
    reportNumber = "EOS-2018-02, TUM-HEP-1172/18",
    doi = "10.1007/JHEP01(2019)150",
    journal = "JHEP",
    volume = "01",
    pages = "150",
    year = "2019"
}

@article{Faller:2008tr,
    author = "Faller, S. and Khodjamirian, A. and Klein, Ch. and Mannel, Th.",
    title = "{$B \to D^{(*)}$ Form Factors from QCD Light-Cone Sum Rules}",
    eprint = "0809.0222",
    archivePrefix = "arXiv",
    primaryClass = "hep-ph",
    reportNumber = "SI-HEP-2008-13",
    doi = "10.1140/epjc/s10052-009-0968-4",
    journal = "Eur. Phys. J. C",
    volume = "60",
    pages = "603--615",
    year = "2009"
}

@article{Khodjamirian:2022vta,
    author = "Khodjamirian, Alexander and Wald, Marcel",
    title = "{B-meson decay into a proton and dark antibaryon from QCD light-cone sum rules}",
    eprint = "2206.11601",
    archivePrefix = "arXiv",
    primaryClass = "hep-ph",
    reportNumber = "SI-HEP-2022-08, P3H-22-060",
    doi = "10.1016/j.physletb.2022.137434",
    journal = "Phys. Lett. B",
    volume = "834",
    pages = "137434",
    year = "2022"
}

@article{Boushmelev:2023huu,
    author = "Boushmelev, Anastasia and Wald, Marcel",
    title = "{Higher twist corrections to B-meson decays into a proton and dark antibaryon from QCD light-cone sum rules}",
    eprint = "2311.13482",
    archivePrefix = "arXiv",
    primaryClass = "hep-ph",
    reportNumber = "SI-HEP-2023-29, P3H-23-093",
    doi = "10.1103/PhysRevD.109.055049",
    journal = "Phys. Rev. D",
    volume = "109",
    number = "5",
    pages = "055049",
    year = "2024"
}

@article{Elor:2022jxy,
    author = "Elor, Gilly and Guerrera, Alfredo Walter Mario",
    title = "{Branching fractions of B meson decays in Mesogenesis}",
    eprint = "2211.10553",
    archivePrefix = "arXiv",
    primaryClass = "hep-ph",
    doi = "10.1007/JHEP02(2023)100",
    journal = "JHEP",
    volume = "02",
    pages = "100",
    year = "2023"
}

@article{Lenz:2024rwi,
    author = {Lenz, Alexander and Mohamed, Ali and W{\"u}thrich, Zachary},
    title = "{Constraining B-mesogenesis models with inclusive and exclusive decays}",
    eprint = "2412.14947",
    archivePrefix = "arXiv",
    primaryClass = "hep-ph",
    reportNumber = "P3H-24-099,SI-HEP-2024-30",
    doi = "10.1007/JHEP08(2025)141",
    journal = "JHEP",
    volume = "08",
    pages = "141",
    year = "2025"
}

@article{Shi:2023riy,
    author = "Shi, Yu-Ji and Xing, Ye and Xing, Zhi-Peng",
    title = "{Semi-inclusive decays of B meson into a dark anti-baryon and baryons}",
    eprint = "2305.17622",
    archivePrefix = "arXiv",
    primaryClass = "hep-ph",
    doi = "10.1140/epjc/s10052-023-11930-z",
    journal = "Eur. Phys. J. C",
    volume = "83",
    number = "8",
    pages = "744",
    year = "2023"
}

@article{ParticleDataGroup:2024cfk,
    author = "Navas, S. and others",
    collaboration = "Particle Data Group",
    title = "{Review of particle physics}",
    doi = "10.1103/PhysRevD.110.030001",
    journal = "Phys. Rev. D",
    volume = "110",
    number = "3",
    pages = "030001",
    year = "2024"
}

@article{Espriu:1983hu,
    author = "Espriu, D. and Pascual, P. and Tarrach, R.",
    title = "{Baryon  Masses and Chiral Symmetry Breaking}",
    doi = "10.1016/0550-3213(83)90663-6",
    journal = "Nucl. Phys. B",
    volume = "214",
    pages = "285--298",
    year = "1983"
}

@article{FlavourLatticeAveragingGroupFLAG:2024oxs,
    author = "Aoki, Y. and others",
    collaboration = "Flavour Lattice Averaging Group (FLAG)",
    title = "{FLAG Review 2024}",
    eprint = "2411.04268",
    archivePrefix = "arXiv",
    primaryClass = "hep-lat",
    reportNumber = "CERN-TH-2024-192, FERMILAB-PUB-24-0785-T",
    month = "11",
    year = "2024"
}

@article{Khodjamirian:2023wol,
    author = "Khodjamirian, Alexander and Meli{\'c}, Bla{\v{z}}enka and Wang, Yu-Ming",
    title = "{A guide to the QCD light-cone sum rules for $b$-quark decays}",
    eprint = "2311.08700",
    archivePrefix = "arXiv",
    primaryClass = "hep-ph",
    reportNumber = "SI-HEP-2023-25, P3H-23-087; RBI-ThPhys-2023-37",
    doi = "10.1140/epjs/s11734-023-01046-6",
    journal = "Eur. Phys. J. ST",
    volume = "233",
    number = "2",
    pages = "271--298",
    year = "2024"
}

@article{Braun:2003wx,
    author = "Braun, V. M. and Ivanov, D. Yu. and Korchemsky, G. P.",
    title = "{The B meson distribution amplitude in QCD}",
    eprint = "hep-ph/0309330",
    archivePrefix = "arXiv",
    reportNumber = "LPT-ORSAY-03-63",
    doi = "10.1103/PhysRevD.69.034014",
    journal = "Phys. Rev. D",
    volume = "69",
    pages = "034014",
    year = "2004"
}

@article{Khodjamirian:2020hob,
    author = "Khodjamirian, Alexander and Mandal, Rusa and Mannel, Thomas",
    title = "{Inverse moment of the B$_{s}$-meson distribution amplitude from QCD sum rule}",
    eprint = "2008.03935",
    archivePrefix = "arXiv",
    primaryClass = "hep-ph",
    reportNumber = "SI-HEP-2020-20, P3H-20-039",
    doi = "10.1007/JHEP10(2020)043",
    journal = "JHEP",
    volume = "10",
    pages = "043",
    year = "2020"
}

@article{Nishikawa:2011qk,
    author = "Nishikawa, Tetsuo and Tanaka, Kazuhiro",
    title = "{QCD Sum Rules for Quark-Gluon Three-Body Components in the B Meson}",
    eprint = "1109.6786",
    archivePrefix = "arXiv",
    primaryClass = "hep-ph",
    reportNumber = "J-PARC-TH-0032",
    doi = "10.1016/j.nuclphysb.2013.12.007",
    journal = "Nucl. Phys. B",
    volume = "879",
    pages = "110--142",
    year = "2014"
}

@article{Rahimi:2020zzo,
    author = "Rahimi, Muslem and Wald, Marcel",
    title = "{QCD sum rules for parameters of the B-meson distribution amplitudes}",
    eprint = "2012.12165",
    archivePrefix = "arXiv",
    primaryClass = "hep-ph",
    reportNumber = "P3H-20-082, SI-HEP-2020-35",
    doi = "10.1103/PhysRevD.104.016027",
    journal = "Phys. Rev. D",
    volume = "104",
    number = "1",
    pages = "016027",
    year = "2021"
}

@article{Anikin:2013aka,
    author = "Anikin, I. V. and Braun, V. M. and Offen, N.",
    title = "{Nucleon Form Factors and Distribution Amplitudes in QCD}",
    eprint = "1310.1375",
    archivePrefix = "arXiv",
    primaryClass = "hep-ph",
    doi = "10.1103/PhysRevD.88.114021",
    journal = "Phys. Rev. D",
    volume = "88",
    pages = "114021",
    year = "2013"
}

@article{Lenz:2009ar,
    author = "Lenz, Alexander and Gockeler, Meinulf and Kaltenbrunner, Thomas and Warkentin, Nikolaus",
    title = "{The Nucleon Distribution Amplitudes and their application to nucleon form factors and the $N \to \Delta$ transition at intermediate values of $Q^2$}",
    eprint = "0903.1723",
    archivePrefix = "arXiv",
    primaryClass = "hep-ph",
    doi = "10.1103/PhysRevD.79.093007",
    journal = "Phys. Rev. D",
    volume = "79",
    pages = "093007",
    year = "2009"
}

@article{Braun:2001tj,
    author = "Braun, Vladimir M. and Lenz, A. and Mahnke, N. and Stein, E.",
    title = "{Light cone sum rules for the nucleon form-factors}",
    eprint = "hep-ph/0112085",
    archivePrefix = "arXiv",
    doi = "10.1103/PhysRevD.65.074011",
    journal = "Phys. Rev. D",
    volume = "65",
    pages = "074011",
    year = "2002"
}

@article{Descotes-Genon:2019bud,
    author = "Descotes-Genon, S{\'e}bastien and Khodjamirian, Alexander and Virto, Javier",
    title = "{Light-cone sum rules for $B\to K\pi$ form factors and applications to rare decays}",
    eprint = "1908.02267",
    archivePrefix = "arXiv",
    primaryClass = "hep-ph",
    reportNumber = "LPT-ORSAY/19-31, SI-HEP-2019-11, TUM-HEP-1161/18, MIT-CTP/5058,
  NIOBE-2019-01",
    doi = "10.1007/JHEP12(2019)083",
    journal = "JHEP",
    volume = "12",
    pages = "083",
    year = "2019"
}

@article{Balitsky:1987bk,
    author = "Balitsky, I. I. and Braun, Vladimir M.",
    title = "{Evolution Equations for QCD String Operators}",
    reportNumber = "LENINGRAD-87-1351",
    doi = "10.1016/0550-3213(89)90168-5",
    journal = "Nucl. Phys. B",
    volume = "311",
    pages = "541--584",
    year = "1989"
}

@article{Bourrely:2008za,
    author = "Bourrely, Claude and Caprini, Irinel and Lellouch, Laurent",
    title = "{Model-independent description of $B \to  \pi \ell \nu$ decays and a determination of |V(ub)|}",
    eprint = "0807.2722",
    archivePrefix = "arXiv",
    primaryClass = "hep-ph",
    reportNumber = "CPT-P36-2007",
    doi = "10.1103/PhysRevD.82.099902",
    journal = "Phys. Rev. D",
    volume = "79",
    pages = "013008",
    year = "2009",
    note = "[Erratum: Phys.Rev.D 82, 099902 (2010)]"
}

@article{Mohamed:2025zgx,
    author = "Mohamed, Ali",
    title = "{Are Subleading Effects Really Subleading? $B$-Meson Decays in Mesogenesis}",
    eprint = "2511.04858",
    archivePrefix = "arXiv",
    primaryClass = "hep-ph",
    reportNumber = "P3H-25-089, SI-HEP-2025-24",
    month = "11",
    year = "2025"
}

@article{BaBar:2023dtq,
    author = "Lees, J. P. and others",
    collaboration = "BaBar",
    title = "{Search for Evidence of Baryogenesis and Dark Matter in $B^+ \to \psi_D+p$ Decays at BABAR}",
    eprint = "2306.08490",
    archivePrefix = "arXiv",
    primaryClass = "hep-ex",
    reportNumber = "BABAR-PUB-23/03 SLAC-PUB-17731",
    doi = "10.1103/PhysRevLett.131.201801",
    journal = "Phys. Rev. Lett.",
    volume = "131",
    number = "20",
    pages = "201801",
    year = "2023"
}

@article{Belle-II:2026tyb,
    author = "Abumusabh, M. and others",
    collaboration = "Belle-II, Belle",
    title = "{A search for feebly-interacting particles in $B$ decays with missing energy at Belle}",
    eprint = "2601.07104",
    archivePrefix = "arXiv",
    primaryClass = "hep-ex",
    reportNumber = "Belle II Preprint: 2025-029 KEK Preprint: 2025-36",
    month = "1",
    year = "2026"
}

@article{BaBar:2023rer,
    author = "Lees, J. P. and others",
    collaboration = "BaBar",
    title = "{Search for B mesogenesis at BaBar}",
    eprint = "2302.00208",
    archivePrefix = "arXiv",
    primaryClass = "hep-ex",
    doi = "10.1103/PhysRevD.107.092001",
    journal = "Phys. Rev. D",
    volume = "107",
    number = "9",
    pages = "092001",
    year = "2023"
}

@article{BaBar:2024qqx,
    author = "Lees, J. P. and others",
    collaboration = "BaBar",
    title = "{Search for baryogenesis and dark matter in B+{\textrightarrow}{\ensuremath{\Lambda}}c++invisible decays}",
    eprint = "2412.06950",
    archivePrefix = "arXiv",
    primaryClass = "hep-ex",
    reportNumber = "BABAR-PUB-24/001, SLAC-PUB-241025",
    doi = "10.1103/PhysRevD.111.L031101",
    journal = "Phys. Rev. D",
    volume = "111",
    number = "3",
    pages = "L031101",
    year = "2025"
}

@article{Belle:2021gmc,
    author = "Hadjivasiliou, C. and others",
    collaboration = "Belle",
    title = "{Search for $B^0$ meson decays into $\Lambda$ and missing energy with a hadronic tagging method at Belle}",
    eprint = "2110.14086",
    archivePrefix = "arXiv",
    primaryClass = "hep-ex",
    reportNumber = "Belle Preprint 2021-24; KEK Preprint 2021-29; PNNL-SA-167279",
    doi = "10.1103/PhysRevD.105.L051101",
    journal = "Phys. Rev. D",
    volume = "105",
    number = "5",
    pages = "L051101",
    year = "2022"
}

@article{Shi:2024uqs,
    author = "Shi, Yu-Ji and Xing, Ye and Xing, Zhi-Peng",
    title = "{Heavy baryon decays into light meson and dark baryon within LCSR}",
    eprint = "2401.14120",
    archivePrefix = "arXiv",
    primaryClass = "hep-ph",
    doi = "10.1140/epjc/s10052-024-12663-3",
    journal = "Eur. Phys. J. C",
    volume = "84",
    number = "3",
    pages = "306",
    year = "2024"
}

@article{Hiller:2026osz,
    author = "Hiller, Gudrun and Rodr{\'\i}guez-S{\'a}nchez, Antonio and Wendler, Daniel",
    title = "{Probing baryon number with missing energy}",
    eprint = "2602.15936",
    archivePrefix = "arXiv",
    primaryClass = "hep-ph",
    month = "2",
    year = "2026"
}

\end{document}